\documentclass[preprint,12pt]{emulateapj}
\usepackage{epsfig}
\usepackage{natbib}
\usepackage{amssymb}
\usepackage{amsbsy}
\usepackage{natbib}
\usepackage[mathcal]{euscript}
\bibpunct{(}{)}{,}{a}{}{,}
\begin{document}
 
%macro for invoking today's date (when TeX is run on your file)
\def\today{\number\year\space \ifcase\month\or  January\or February\or
        March\or April\or May\or June\or July\or August\or
September\or
        October\or November\or December\fi\space \number\day}
%\fraction makes a nice fraction
\def\fraction#1/#2{\leavevmode\kern.1em
 \raise.5ex\hbox{\the\scriptfont0 #1}\kern-.1em
 /\kern-.15em\lower.25ex\hbox{\the\scriptfont0 #2}}
\def\simlt{\vcenter{\hbox{$<$}\offinterlineskip\hbox{$\sim$}}}
\def\simgt{\vcenter{\hbox{$>$}\offinterlineskip\hbox{$\sim$}}}
%\simlt and \simgt produce > and < signs with twiddle underneath
\def\spose#1{\hbox to 0pt{#1\hss}}
\def\etal{et al.\ }

\renewcommand{\thefootnote}{\fnsymbol{footnote}}

\title{Precise high-cadence time series observations of five variable young stars in Auriga with {\em
MOST}\footnotemark[*]}
\author{Ann Marie Cody\altaffilmark{1,2}, Jamie Tayar\altaffilmark{1,3}, Lynne A. Hillenbrand\altaffilmark{1},
Jaymie M. Matthews\altaffilmark{4}, and Thomas Kallinger\altaffilmark{5}}
\email{amc@ipac.caltech.edu}

\altaffiltext{1}{California Institute of Technology, Department of Astrophysics, MC 249-17, Pasadena, CA 91125, USA}
\altaffiltext{2}{Spitzer Science Center, California Institute of Technology, Pasadena, CA 91125, USA}
\altaffiltext{3}{Department of Astronomy, The Ohio State University, 140 W 18th Ave, Columbus, OH 43210, USA}
\altaffiltext{4}{Department of Physics \& Astronomy, University of British Columbia, 6224 Agricultural Road, Vancouver,
British Columbia V6T 1Z1 Canada}
\altaffiltext{5}{Institut f\"{u}r Astronomie, Universit\"{a}t Wien, T\"{u}rkenschanzstrasse 17, A-1180 Vienna, Austria}

\footnotetext[*]{Based on data from the {\em MOST} satellite, a Canadian Space Agency mission, 
jointly operated by Systems Canada Inc. (MSCI), formerly part of Dynacon, Inc., the University of 
Toronto Institute for Aerospace Studies, and the University of British Columbia with the assistance of the University 
of Vienna.}

\begin{abstract}
To explore young star variability on a large range of timescales, we have used the {\em MOST} satellite to obtain 24 
days of continuous, sub-minute cadence, high-precision optical photometry on a field of classical and weak-lined T~Tauri 
stars (TTS) in the Taurus-Auriga star formation complex.  Observations of AB Aurigae, SU Aurigae, V396 Aurigae, V397 
Aurigae, and HD 31305 reveal brightness fluctuations at the 1--10\% level on timescales of hours to weeks. We have 
further assessed the variability properties with Fourier, wavelet, and autocorrelation techniques, identifying one 
significant period per star. We present spot models in an attempt to fit the periodicities, but find that we cannot 
fully account for the observed variability. Rather, all stars exhibit a mixture of periodic and aperiodic behavior, with 
the latter dominating stochastically on timescales less than several days. After removal of the main periodicity, 
periodograms for each light curve display power law trends consistent with those seen for other young accreting stars. 
Several of our targets exhibited unusual variability patterns not anticipated by prior studies, and we propose that this 
behavior originates with the circumstellar disks. The {\em MOST} observations underscore the need for investigation of 
TTS light variations on a wide range of timescales in order to elucidate the physical processes responsible; we provide 
guidelines for future time series observations.

\end{abstract}


\section{Introduction}

The widespread photometric activity seen in almost all young stellar objects is a defining characteristic of the 
class \citep{1945ApJ...102..168J,1962AdA&A...1...47H}. In particular, classical T Tauri stars (CTTS) display 
brightness variations from the millimagnitude \citep{Cody10,Cody11} to magnitude level 
\citep{Herbst02,Carpenter01,Carpenter02}, on timescales from days to years. While some CTTS light curves appear to 
contain regular sinusoidal patterns, the dominant form of the variations is aperiodic, with at times abrupt and 
unpredictable changes. The photometric variation of weak-lined T Tauri stars (WTTS; so called because their 
signatures of accretion are weak), on the other hand, appears to be dominated by spot rotation signatures, often 
with additional low level stochastic fluctuations. These diverse properties have been variously attributed to 
rotational modulation of cool magnetic spots, enhanced chromospheric activity, hot spots from columns of 
magnetically channeled shocked inflowing gas, unsteady mass accretion, and occultation or shadowing by material in a 
surrounding circumstellar disk. Combinations of photometric, spectroscopic and polarimetric techniques have 
illuminated some of these possibilities, but a rigorous explanation of the erratic, often non-periodic nature of 
brightness fluctuations awaits.

Central to the problem of understanding photometric activity in T Tauri stars is the need for continuous 
monitoring on many different timescales. Photometry from ground-based facilities is limited in both 
precision and time coverage, consequently making efforts to model the light curves difficult, if not 
impossible. Routine gaps in observation are acceptable for the detection of periodic brightness 
variations, since these signals generally can be isolated in the frequency domain. But the flickering 
portions of T~Tauri lightcurves are not amenable to Fourier and similar analysis methods.

To date only a few studies have monitored young stars at high precision and cadence over long, continuous 
time baselines. Among these, \citet{2010A&A...519A..88A} presented uninterrupted 23-day lightcurves of 
NGC~2264 members with 0.5--5 mmag precision from the {\em CoRoT} mission. They 
specifically identified objects with light curves resembling that of AA~Tau, a CTTS with variability 
attributed to the repeated passage of warped disk material in front of the stellar photosphere every 
few days. The {\em CoRoT} dataset illustrated that AA~Tau-like fading events are common in young stars with 
infrared excess and include not only dramatic brightness decreases but also erratic lower amplitude 
fluctuations indicative of additional dynamics in the inner disk. The full complexity of this behavior 
was previously impossible to capture with ground-based time series.

\citet{2010A&A...522A.113R} used the Microvariability and Oscillations of STars telescope \citep[hereafter {\em 
MOST};][]{2003PASP..115.1023W,2004Natur.430...51M} to observe the Herbig Ae star HD~37806 for 21 days at $\sim$3 mmag 
precision, combining this dataset with nine seasons of observations by the All Sky Automated Survey (ASAS). They 
characterized the light curve as weakly periodic on a timescale of $\sim$1.5 days, and otherwise stochastic, with 
flares or accretion instabilities comprising amplitudes from 0.03\% to 5\% on timescales of minutes to years. Another 
Herbig Ae star, HD~142666, was shown by \citet{2009A&A...494.1031Z} to exhibit both $\delta$~Scuti pulsations and 
irregular UX~Ori type variations attributed to the circumstellar disk.

Additional high cadence investigation of young star photometric behavior spanning multiple week timescales was 
presented by \citet{2008MNRAS.391.1913R} and \citet{2011MNRAS.415.1119S,2011MNRAS.410.2725S}. They acquired photometry 
at the 1--7 mmag precision level on TW Hya (nearly continuous observations of 40 and 46 days) and five other T~Tauri 
stars in the Taurus-Auriga and Lupus star forming regions (run durations of 12 or 21 days) from {\em MOST}. For the 
WTTS of their sample, most of the observed variability is well modeled by a collection of differentially rotating 
surface spots. The CTTS light curves present variability that is more challenging to interpret. In the case of RY~Tau, 
there are two pronounced brightness dips of $\sim$0.2 mag depth as well as lower amplitude, transient 
oscillatory behavior superimposed on a longer timescale trend. For TW Hya, on the other hand, flicker noise behavior (power 
proportional to inverse frequency) dominates \citep{2008MNRAS.391.1913R}, although semi-periodic 
features were also observed to form and drift to shorter timescales over the course of a 40-day run 
\citep{2011MNRAS.410.2725S}. \citet{2011MNRAS.410.2725S} attribute this type of variability to the magnetospheric 
accretion process, and in particular instabilities driving the flow of plasma from the inner disk.

This small collection of space-based time series data has highlighted the complexity of young star 
variability and underscored the need for further datasets to determine the nature of the 
irregular and low-amplitude flux variations in these objects. The extent to which they are 
representative of young star light curves in general remains unclear. Currently the YSOVAR project 
\citep{2011ApJ...733...50M,2011ASPC..448....5R} is exploring the yield from multiwavelength precision photometry in 
several young clusters including the Orion Nebula Cluster and NGC~2264.

We have taken advantage of the unique capabilities of the {\em MOST} mission to acquire a further 
high-precision, high-cadence, nearly continuous data stream over a 24-day observing period. Our aim was to 
monitor a handful of erratically varying young stars to determine their short-timescale photometric patterns,
decipher the mix of periodic and aperiodic phenomena in operation, and to quantify mathematically 
the aperiodic behavior.  Presented here is {\em MOST} data and analysis of the light curves of 
four T~Tauri and Herbig Ae stars (SU Auriga, AB Auriga, V396~Aur, and V397~Aur), 
and one new candidate early type Taurus member (HD~31305).

The {\em MOST} observations are presented in Section 2 and the variability characterized in section 3.
Section 4 contains discussion of the individual objects, and Section 5 a general discussion of the
implications of the photometric timescales and amplitudes seen in our {\em MOST} 24-day observing sequence. 

\begin{deluxetable*}{ccccccc}
\tabletypesize{\scriptsize}
\tablecolumns{10}
\tablewidth{0pt}
\tablecaption{\bf Targets and Basic Data}
\tablehead{
\colhead{Star}  & \colhead{Other identifiers} & \colhead{$B$} & \colhead{V} & \colhead{SpT} & \colhead{Noise RMS}
& \colhead{Empirical RMS}\\
}
\startdata
AB Aur& HD 31293, HBC 78, 2MASS J04554582+3033043 & 7.1& 7.1& A0/B9& 0.0017 & 0.030\\
HD 31305& 2MASS J04554822+3020165 & 7.7& 7.6 & A0 & 0.00092 & 0.009 \\
SU Aur& HD 282624, HBC 79, 2MASS J04555938+3034015 & 10.2& 9.4 & G2 & 0.0025 & 0.110 \\
V396 Aur & LkCa 19, HBC 426, TAP 56, 2MASS J04553695+3017553  & 12.3 & 11.2 & K0 & 0.019 & 0.050 \\
V397 Aur& HBC 427, TAP 57NW, 2MASS J04560201+3021037 & 12.9& 11.6 & K7$^1$ & 0.024 & 0.058 \\
\enddata
\tablecomments{Basic data on the target stars, as reported by the SIMBAD database and measured from our time
series data. Spectral types are from SIMBAD, except for HD~31305 \citep{2007A&A...468..477A}.
Magnitudes are not simultaneous with our {\em MOST} time series observations.
Noise RMS refers to the standard deviations of our {\em MOST} light curves after subtraction of smoothed median
trends, whereas the empirical RMS is the standard deviation inclusive of intrinsic variability. $^1$V397~Aur is a
single-lined spectroscopic binary system \citep{1988AJ.....96..297W}; the spectral
type corresponds to the primary. A direct imaging study determined $\Delta K$=0.87 for this system
\citep{2011ApJ...731....8K}.}
\end{deluxetable*}

\section{Observations}

In operation since 2003, the 15~cm {\em MOST} telescope produces 
ultra-high precision differential photometry by virtue of its location 820~km above Earth and specially 
designed imaging modes \citep{2006MmSAI..77..282R,2009CoAst.158..162K}. Since the satellite executes a polar orbit 
with period 101.413 minutes, it enables continuous viewing of targets for up to eight weeks in a zone 
(the ``CVZ'') from -18$\arcdeg$ to +36$\arcdeg$ declination. We selected a target field in the 
$\sim$3~Myr Taurus-Auriga star-forming complex based on its location within the CVZ as well as the 
proximity of five suitably bright young targets: SU~Auriga, AB~Auriga, V396~Aur, V397~Auriga, and HD~31305 
(see Table~1). The former four are known young stars in the Taurus-Auriga 
association, while the latter is a newly suggested member based on this work.

Observations took place over 24 days from 2009 December 14 to 2010 January 07. All {\em MOST} images are 
acquired through a broadband filter with wavelength range 350--750 nm. Our three brightest targets (SU~Aur, 
AB~Aur, and HD~31305) were monitored in the direct imaging mode \citep{2006MmSAI..77..282R}, which involves defocusing 
stars to a FWHM of 2--2.5 pixels. It can be applied to up to 10 stars in the magnitude range $V$=6--11, and in 
our case produced photometry with point-to-point precision of 0.001--0.002 mag. For the two fainter 
targets (V396 and V397 Aur), guide star imaging \citep{2005ApJ...635L..77W} offered the best performance, with 
precisions of 0.01--0.02 mag. A third mode, Fabry imaging \citep{2006MNRAS.367.1417R}, produces photometry at 
the $10^{-4}$ mag level but can handle only one target at a time and requires stars brighter than those 
in our sample.

To mitigate pointing effects, {\em MOST} acquires and stacks many ``subexposures'' from 0.3--3 seconds. Total 
exposure times for the combined images was approximately 30--40 seconds, with little dead time between; the 
resulting cadence was 43 seconds. Lightcurves were generated by the {\em MOST} photometric pipeline 
\citep{2006MmSAI..77..282R,2008CoAst.156...48H}, which rejects cosmic
rays and other artifacts, aligns the point spread functions (PSFs) for 
image stacking, and decorrelates background trends from target pixel fluxes. The latter stage was complicated by the 
high-amplitude variability present in most of our targets and hence required manual removal of clear outliers
identified by eye.

Following these reductions, there remained a residual flux contribution from scattered Earthshine 
\citep[e.g.,][]{2006MNRAS.367.1417R,2006MmSAI..77..282R} which introduced high levels of noise over a portion 
($\sim$60\%) of each satellite orbit. The data from these sections was unusable and hence not included in the final 
light curves. The remaining flux points were modulated by a small ($<$0.5\%), non-sinusoidal stray light contribution 
at the 14.2 cycles/day (165 microHz) orbital frequency. In addition to the once-per-orbit data gaps, there are also 
15 larger gaps caused by observations of a source unrelated to our program. The resulting duty cycle is $\sim$35\%.

Since our {\em MOST} data are not amenable to absolute calibration, all photometry presented herein is relative. 
The resulting differential light curves all exhibit variability, such that the RMS values are not reflective of the 
underlying white noise. We have estimated the point-to-point photometric precisions by subtracting out smoothed 
median trends and fitting Gaussian profiles to the remaining noise distributions. The estimated precisions, along with 
the actual RMS values (before subtraction of the median trend), are listed in Table~1.

\section{Variability Characterization and Features}

The final {\em MOST} light curves are presented in Fig.\ 1.
Each of the five targets displayed variability well above the photometric noise level of a few mmag (Table 1). 
Brightness levels fluctuated by at least 0.03 mag (HD~31305) and up to 0.5 mag (SU~Aur)
over the course of the 24-day observations (see Table~1). 
In addition to the range of amplitudes, flux variation is present on a range of timescales. 
Periodic behavior predominates with characteristic timescales of a few days. 
Each star in our sample showed semi-periodic fluctuations. 
Further, a mixture of periodic and aperiodic components is evident in
the light curves, 
in some cases with behavior that is not constant over the entirety of
the light curve. 
For example, the last six days of the SU~Aur light curve show 
a deep fading event; similar episodes have been reported in past literature \citep[e.g.,][]{2003ApJ...590..357D}.
Several objects display transient events, such as flaring. Fig.\ 2 illustrates this in two cases for the light 
curves of SU~Aur and V396~Aur. 

In Section 3.1 we assess the periodic behavior using Fourier analysis. We then examine the evidence for correlated 
aperiodic behavior using wavelet analysis (Section 3.2) and autocorrelation analysis (Section 3.3), which also affirms for 
the periodic objects the results of the Fourier analysis.

\begin{figure}[!h]
\begin{center}
\includegraphics[scale=0.47]{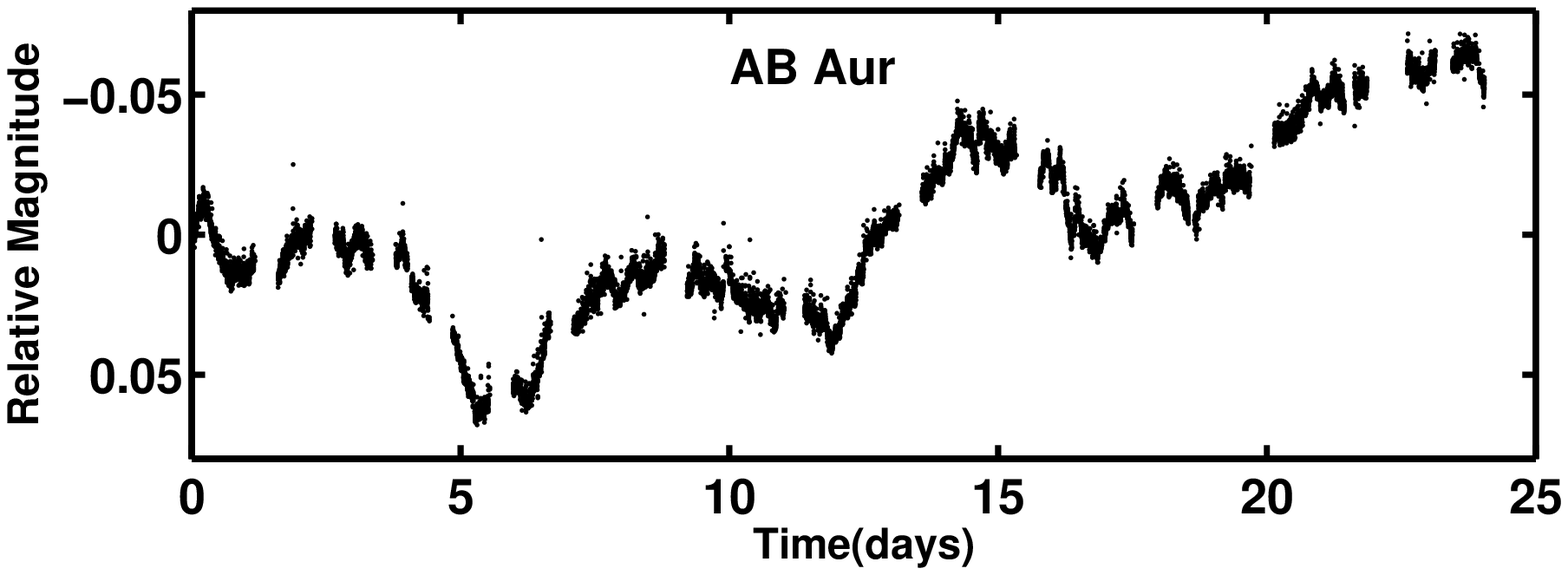}
\includegraphics[scale=0.47]{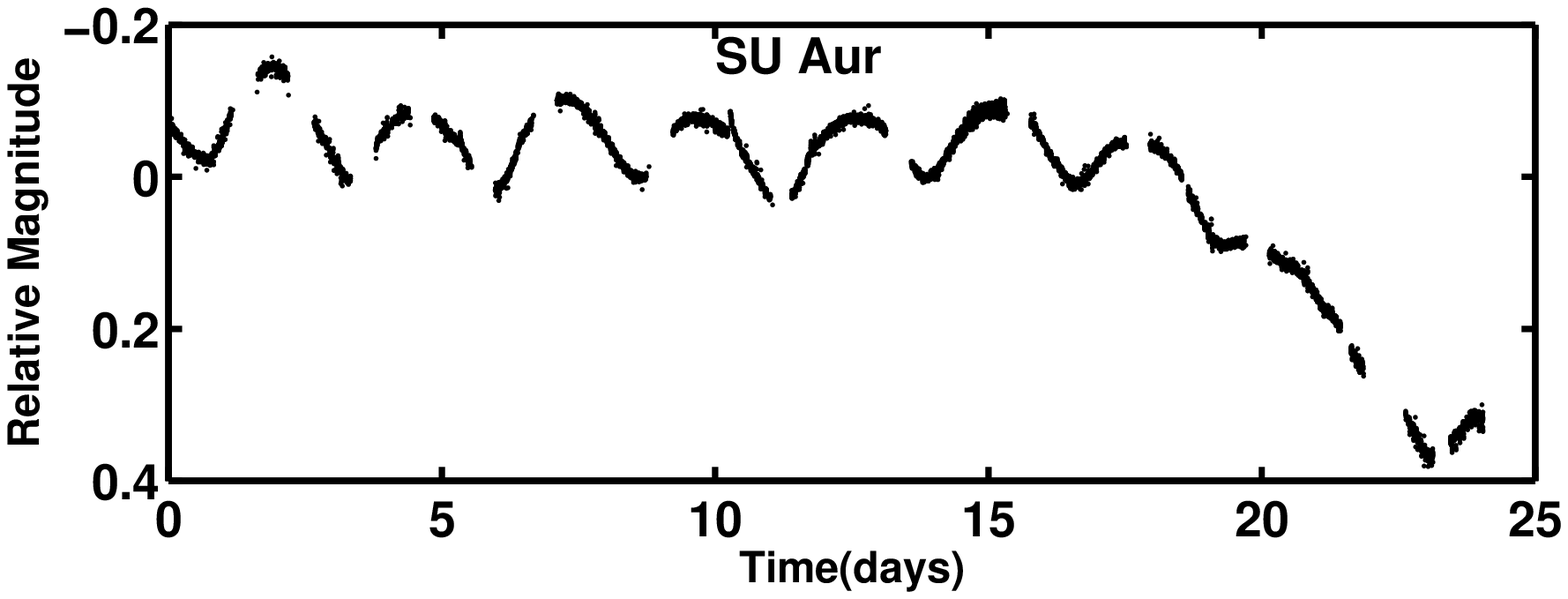}
\includegraphics[scale=0.47]{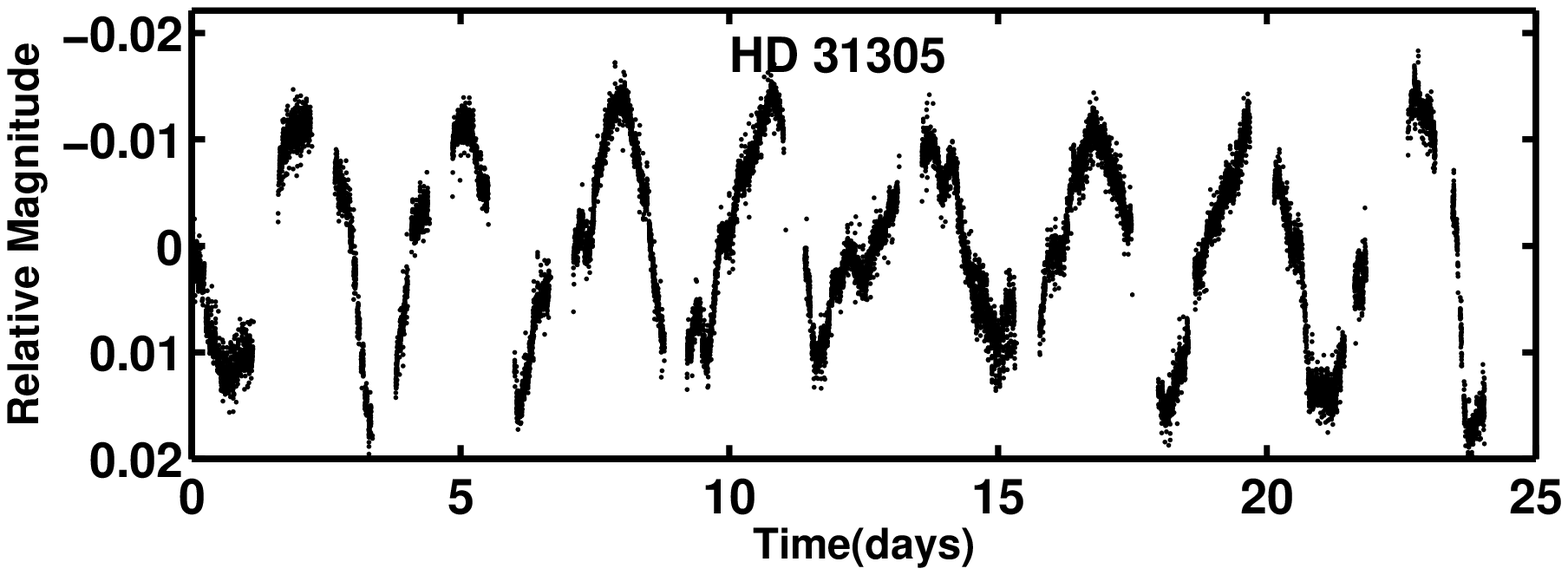}
\includegraphics[scale=0.47]{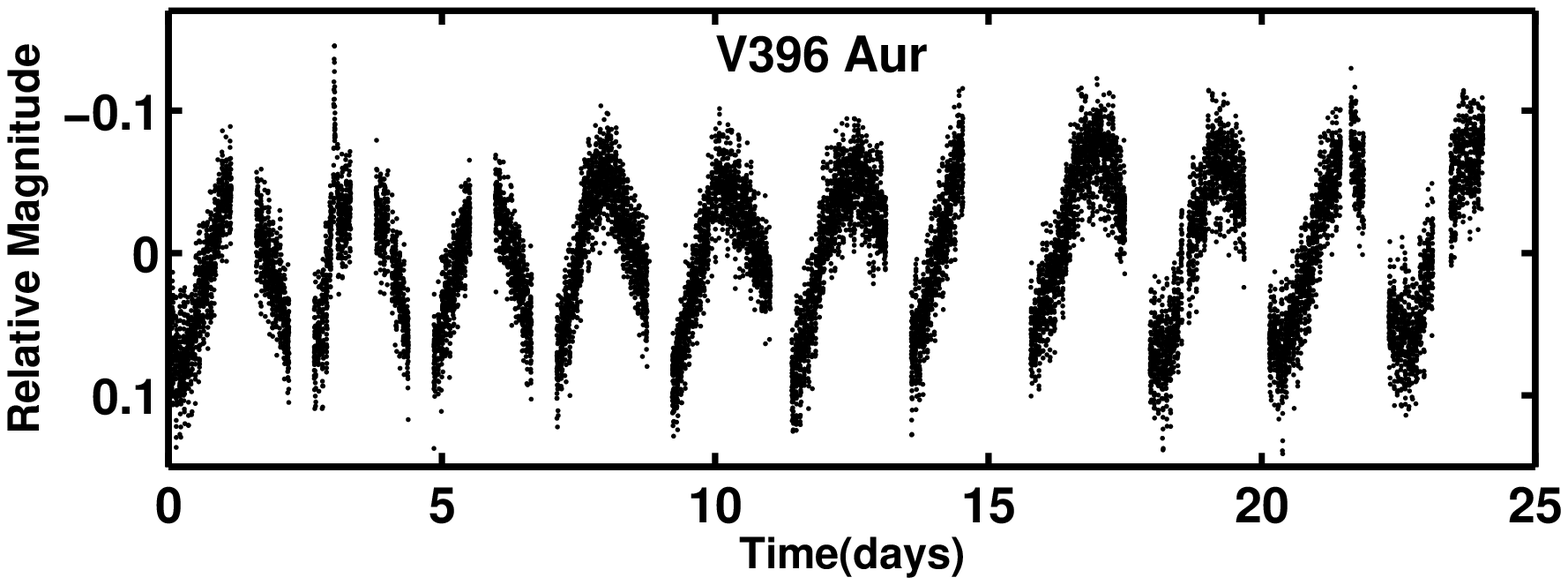}
\includegraphics[scale=0.47]{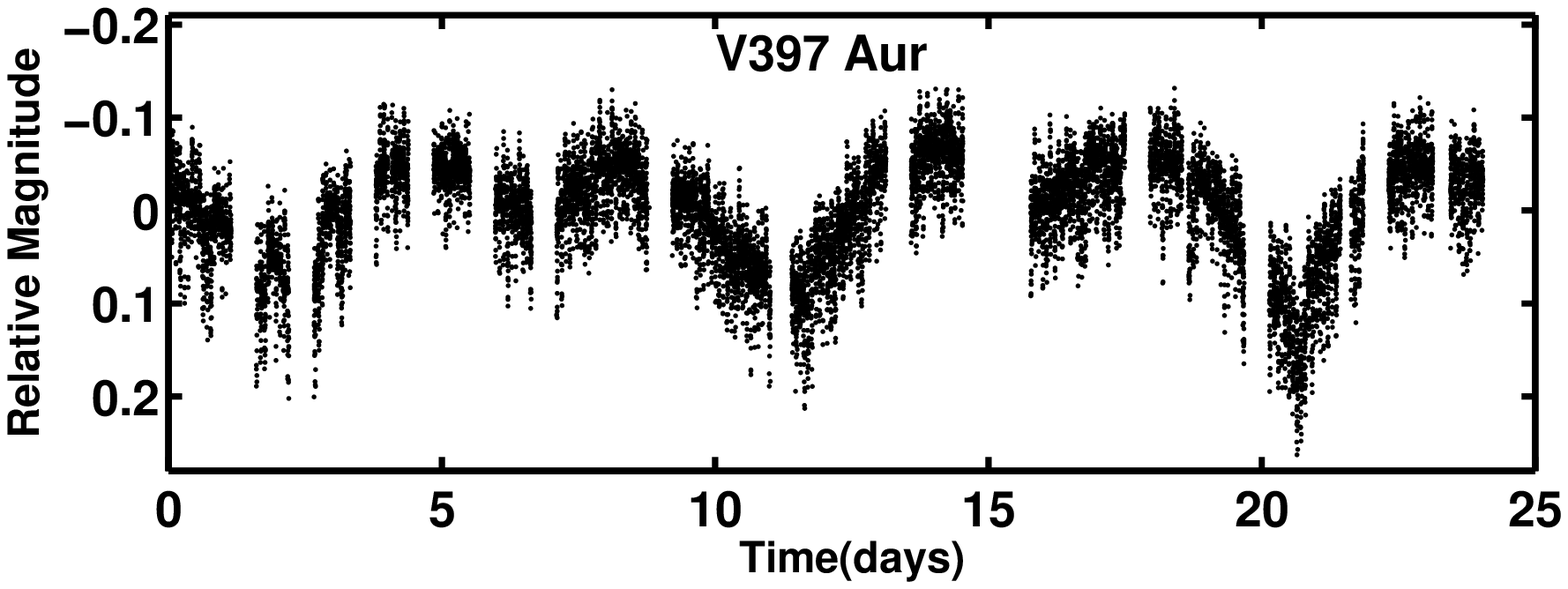}
\end{center}
\caption[]{Differential light curves from {\em MOST} for the five targets, after removal of scattered 
Earthshine. The start of the observations (time=0) corresponds to Julian date 2455180.45.}
\label{} 
\end{figure}

\begin{figure*}[!t]
\begin{center}
\epsscale{0.8}
\plottwo{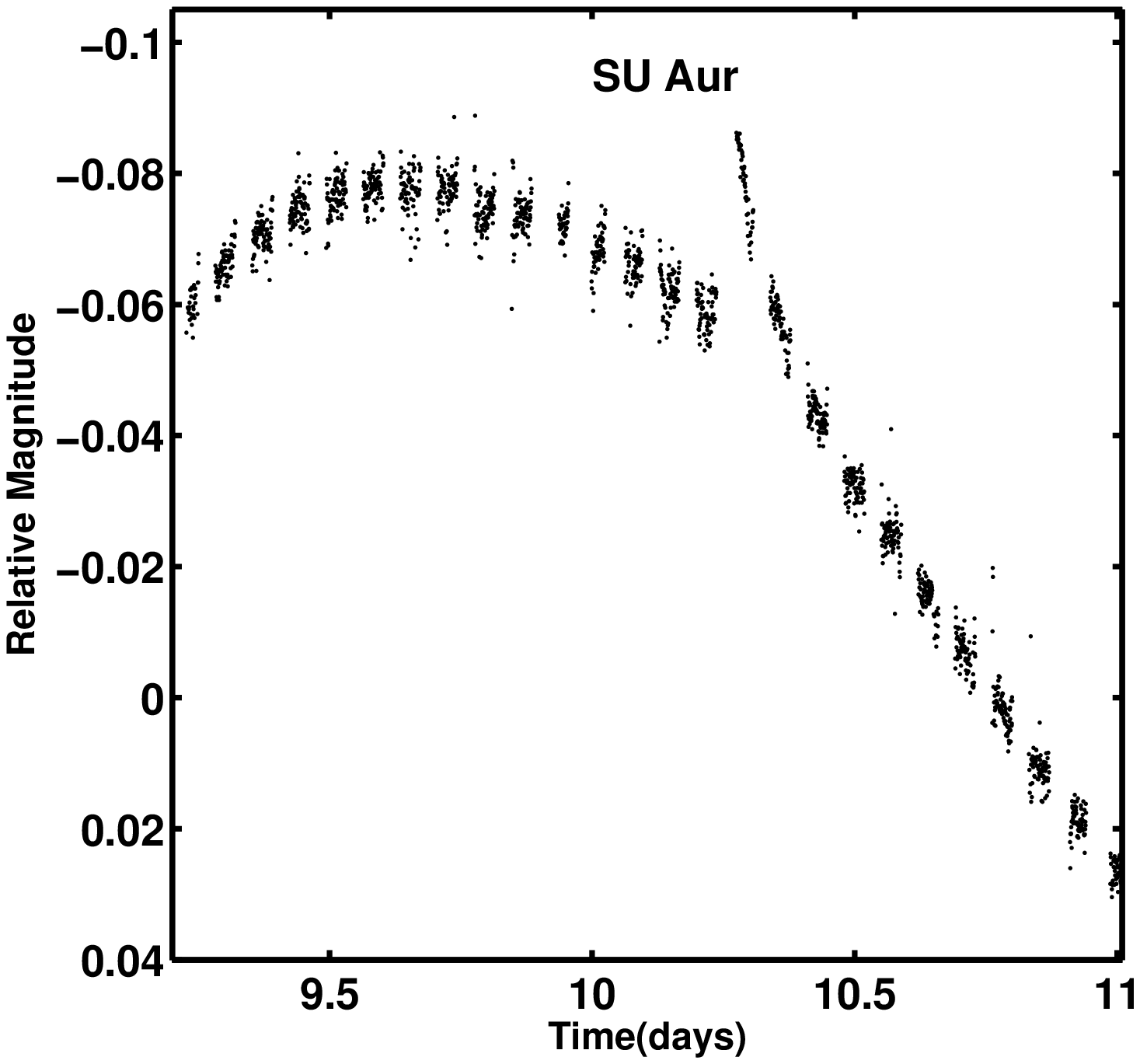}{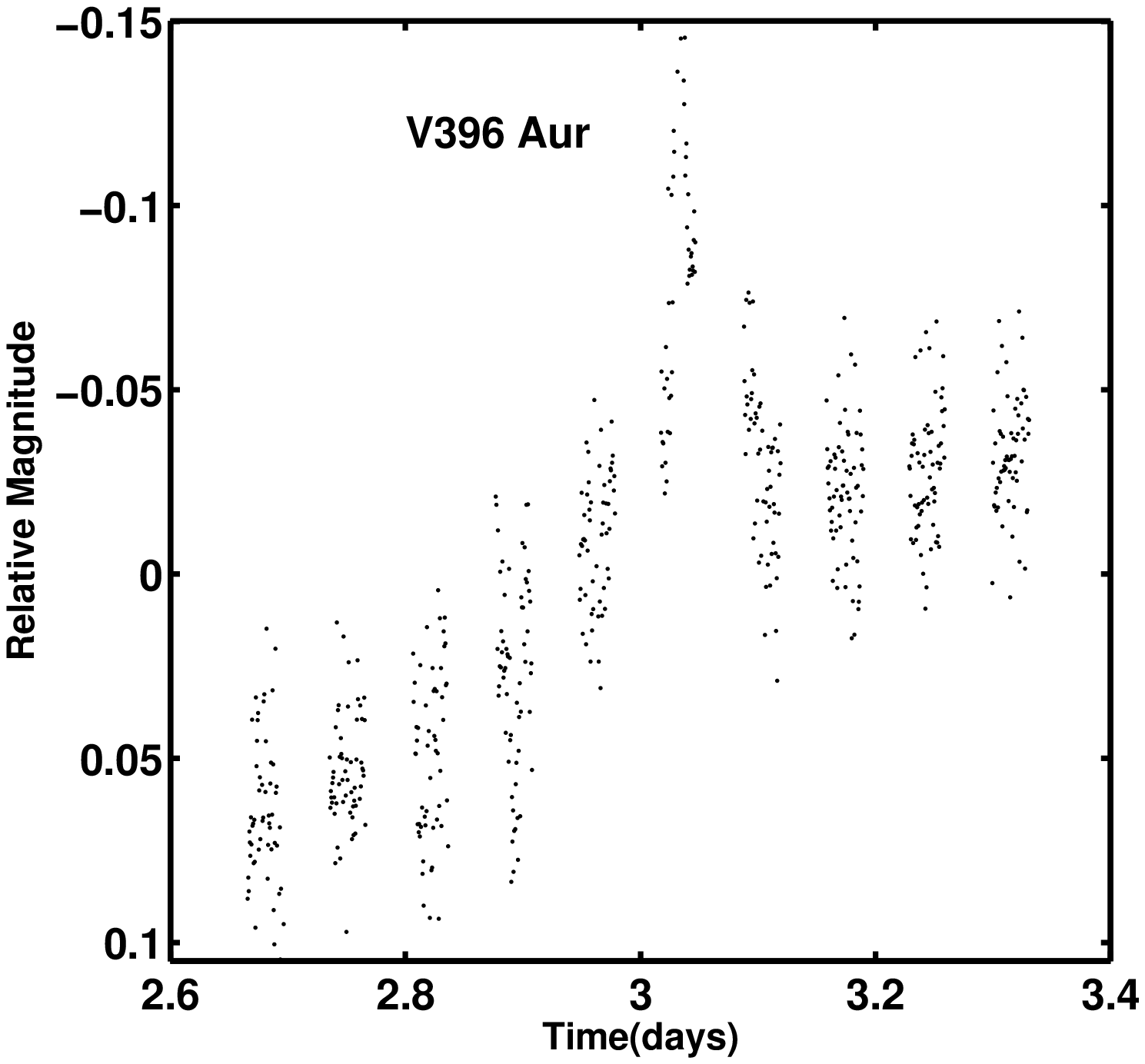}
\end{center}
\caption[]{Flares in two objects occurred during the {\em MOST} observations.}
\label{}
\end{figure*}

\subsection{Fourier analysis}

\begin{deluxetable*}{cccccc}
\tabletypesize{\scriptsize}
\tablecolumns{10}
\tablewidth{0pt}
\tablecaption{\bf Detected periodicities}
\tablehead{\colhead{Star}  & \colhead{Periodogram} & \colhead{Periodogram} & \colhead{Wavelet} &
\colhead{Autocorrelation} &
\colhead{Previous values}
\\
\colhead{} & \colhead{period (d)} & \colhead{amplitude (mag)} & \colhead{period (d)} & \colhead{period (d)} &
\colhead{(d)} \\
}
\startdata
AB Aur& 6.551$\pm$0.015 & 0.0167$\pm$0.0003 & 5.97--7.55 & 5.50$^{+0.24}_{-0.36}$ & 0.54$^1$, 1.38$^2$, 1.83$^2$, 1.76$^3$, 1.33$^4$ \\
HD31305& 2.9379$\pm$0.0009 & 0.01056$\pm$0.00004 & 2.81--3.04 & 2.88$^{+0.03}_{-0.08}$ & \\
SU Aur& 2.661$\pm$0.001 & 0.0456$\pm$0.0002 & 2.39--2.64& 2.64$\pm$0.11& 16$^5$, 3$^6$, 1.7$^{7}$,1.55$^{8}$,2.73$^{8}$,2.7$^{9}$,0.46$^{10}$ \\
V396 Aur& 2.2393$\pm$0.0005 & 0.0652$\pm$0.0003 & 2.21--2.29 & 2.24$^{+0.05}_{-0.04}$ & 2.23$^{11}$, 2.24$^{12}$\\
V397 Aur& 9.54$\pm$0.02& 0.04312$\pm$0.0004& 9.36--9.50 & 9.36$^{+0.03}_{-0.13}$ & 9.32$^{13}$,9.39$^{14}$,10.1$^{15}$ \\
        & 4.6022$\pm$0.0002& 0.0477$\pm$0.0005 &  4.60--4.96& 4.44$^{+0.19}_{-0.02}$ & 4.7$^{16}$
\enddata
\tablecomments{Periods, derived by identifying the largest periodogram peak and its 1-$\sigma$ uncertainties from
Monte Carlo simulations (column 3), by identifying peaks in the wavelet function (column 4; range denotes the extent of the peak),
and by selecting the first
peak in the autocorrelation function and the timescales for which it drops by 1\% from the peak value (column 5).
The two rows under V397~Aur correspond to its individual binary components. We list previously reported rotation periods for
comparison.  References are as follows: $^1$\citet{2005ApJ...622L.133C}, $^2$\citet{1999A&A...345..884C},
$^3$\citet{2007A&A...468..541T}, $^4$\citet{1986ApJ...303..311P}, $^5$\citet{2006PASP..118.1390P},
$^6$\citet{1995ApJ...449..341J}, $^{7}$\citet{2003ApJ...590..357D}, $^{8}$\citet{1987AJ.....94..137H},
$^{9}$\citet{2004MNRAS.348.1301U}, $^{10}$\citet{2007A&A...471..951F}, $^{11}$\citet{2007A&A...461..183G},
$^{12}$\citet{1993A&A...272..176B}, $^{13}$\citet{1993IBVS.3898....1Z},$^{14}$\citet{2007A&A...467..785N},
$^{15}$\citet{1995A&A...299...89B},$^{16}$\citet{1993A&AS..101..485B}}
\end{deluxetable*}

Because the light curve variations appeared to contain periodic components, we used Fourier analysis to identify 
frequencies and assess their significance. Among the unique features of this data were the long baseline and high 
cadence at which it was taken. The nearly uninterrupted sequence of {\em MOST} observations provides an 
opportunity to search for signals in the periodogram without interference from aliasing in the 1--10 day range 
typical of young star rotation and associated spot modulation variability \citep{2009IAUS..258..363I}. Over the 
24-day duration, data points were generally spaced less than one minute apart. Additionally, the instrumental 
error on the brightness values was on the order of one millimagnitude for the data on AB Aurigae, SU Aurigae and 
HD 31305 and slightly higher for V396 and V397~Aur. The high quality of data allow us to search for periodic 
variations ranging from multi-day rotation signatures down to minute-timescale brightness fluctuations, several 
mechanisms for which were proposed recently \citep{2008MNRAS.388..357K,2010A&A...510A..710}.

\begin{figure}[!h]
\begin{center}
\includegraphics[scale=0.45]{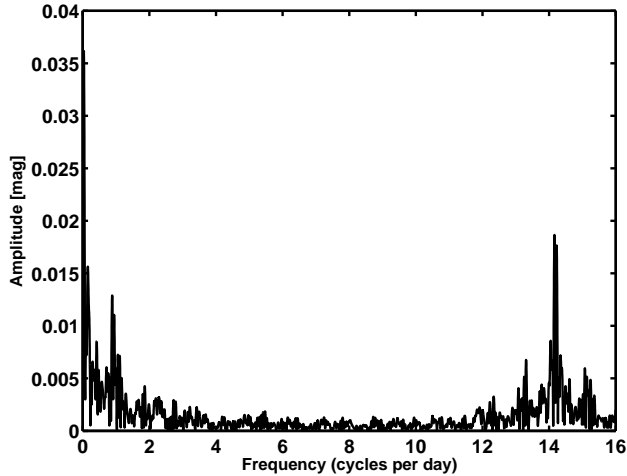}
\end{center}
\caption[]{\label{alias}A portion of the periodogram of AB Aur, demonstrating the
  aliasing pattern at 14.2 d$^{-1}$ that is associated with the {\em MOST}
  telescope's orbital 101-minute period. Power from intrinsic stellar variability
  in the 0--3 d$^{-1}$ range leaks into regions of the periodogram
  with frequencies that are multiples of 14.2. }
\end{figure}

Our selected tool for the analysis of periodic variability is the discrete Fourier transform 
\citep{1975Ap&SS..36..137D} periodogram. The {\em MOST} data sampling pattern includes a built-in periodicity as a 
result of the regular gap in observation (due to high levels of stray light; see Section 2) during the second 
half of each 101-minute orbit. Therefore, the periodogram of any non-constant brightness object contains an 
alias at the associated 14.2 cycles per day (d$^{-1}$) frequency. We display an example of this phenomenon in 
Fig.\ \ref{alias}, where a strong peak at $f$=14.2 is visible in the periodogram of AB~Aur. Because this peak 
repeats at 28.4 and other multiples of 14.2, we restrict our initial period analysis to the frequency range from 0 
to 10 d$^{-1}$ (i.e., periods greater than 2.4 hours). The Nyquist limit, or maximum independent frequency, 
is $\sim$1000 d$^{-1}$ (86 second period).

\begin{figure}[!h]
\begin{center}
\includegraphics[scale=0.47]{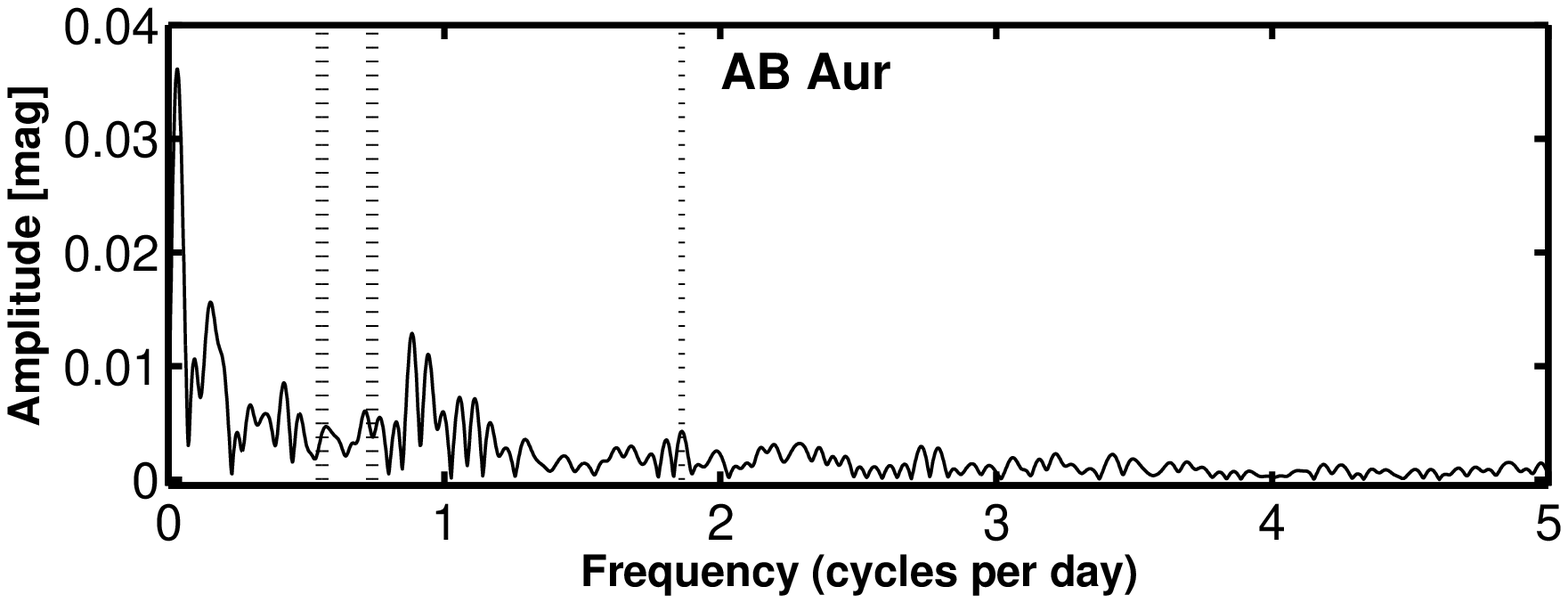}
\includegraphics[scale=0.47]{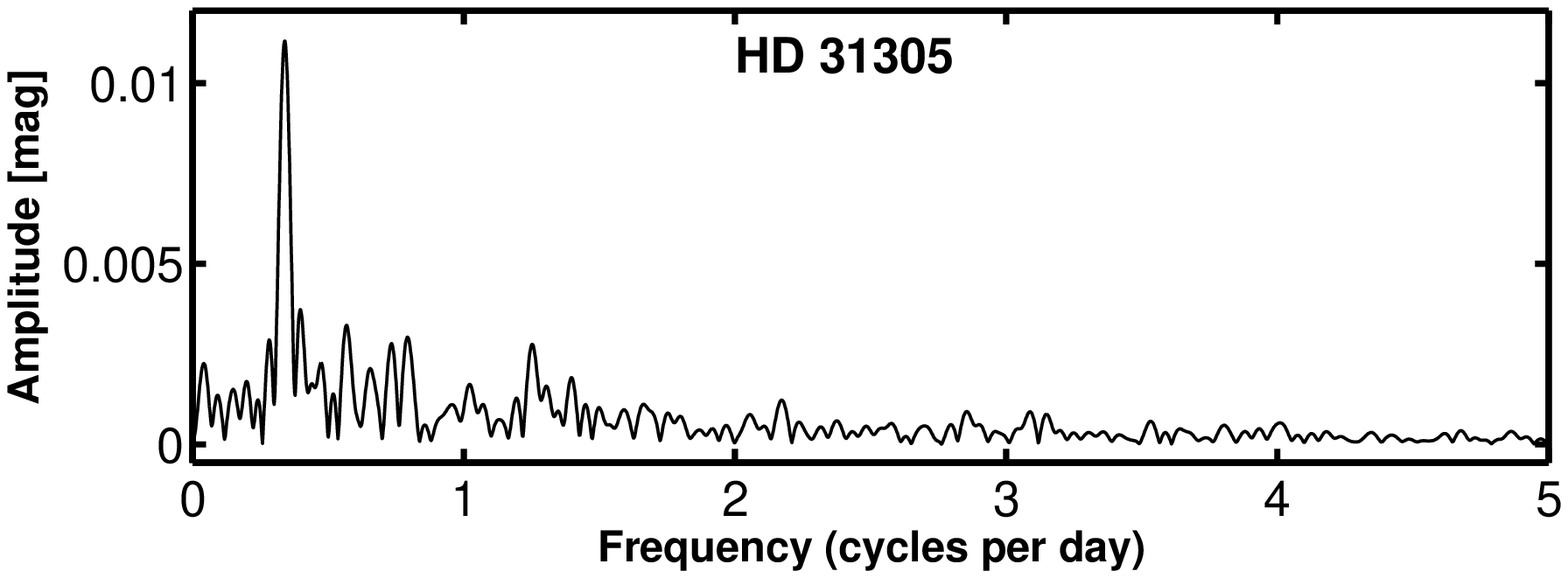}
\includegraphics[scale=0.47]{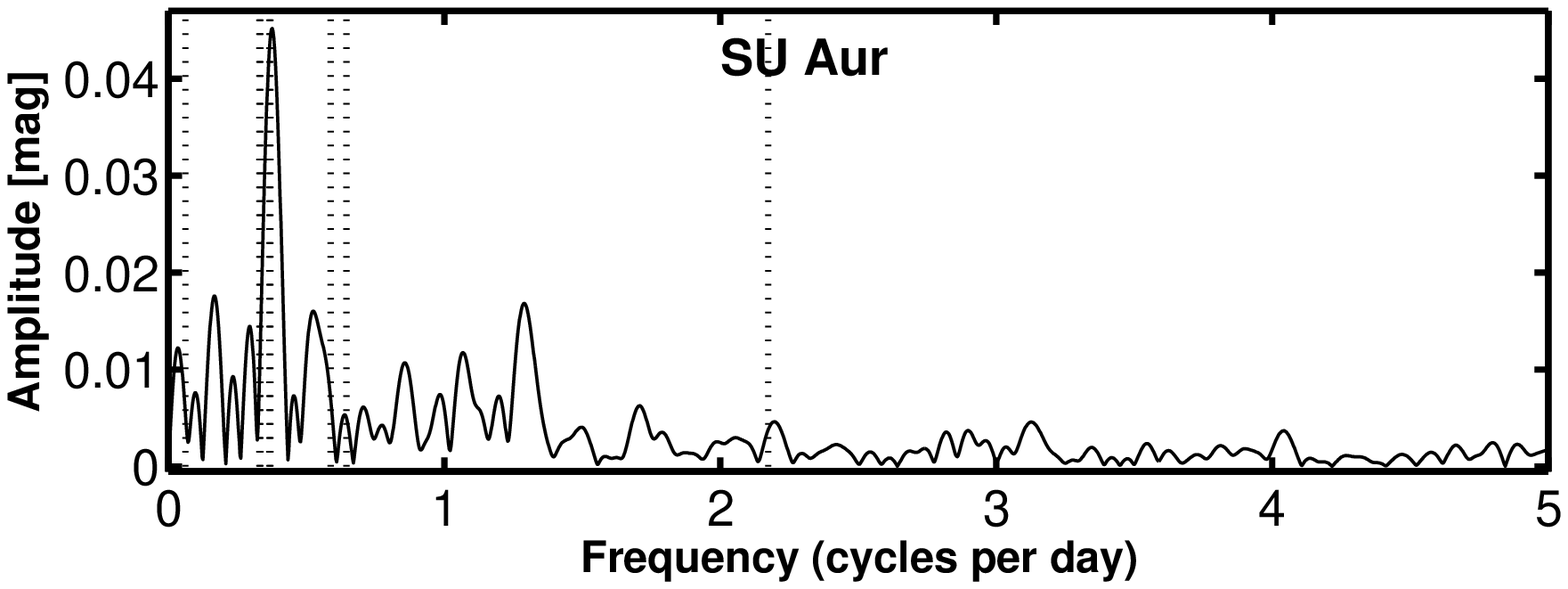}
\includegraphics[scale=0.47]{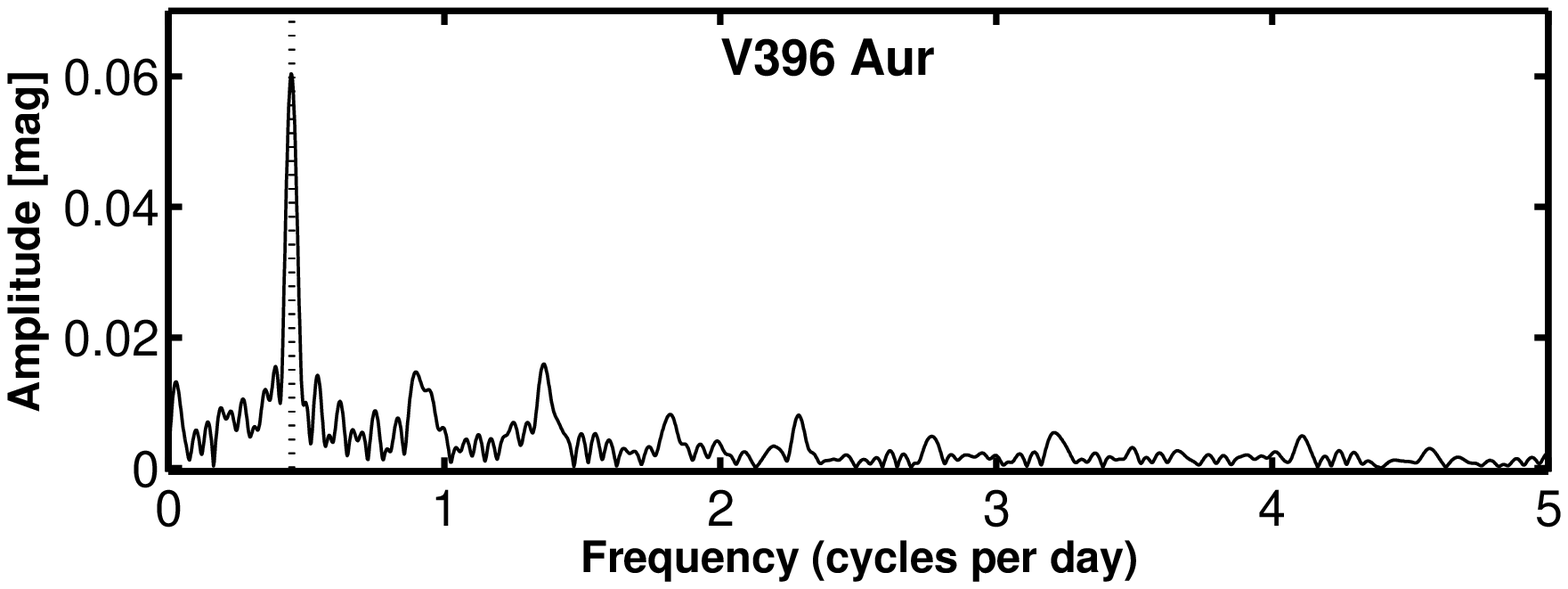}
\includegraphics[scale=0.47]{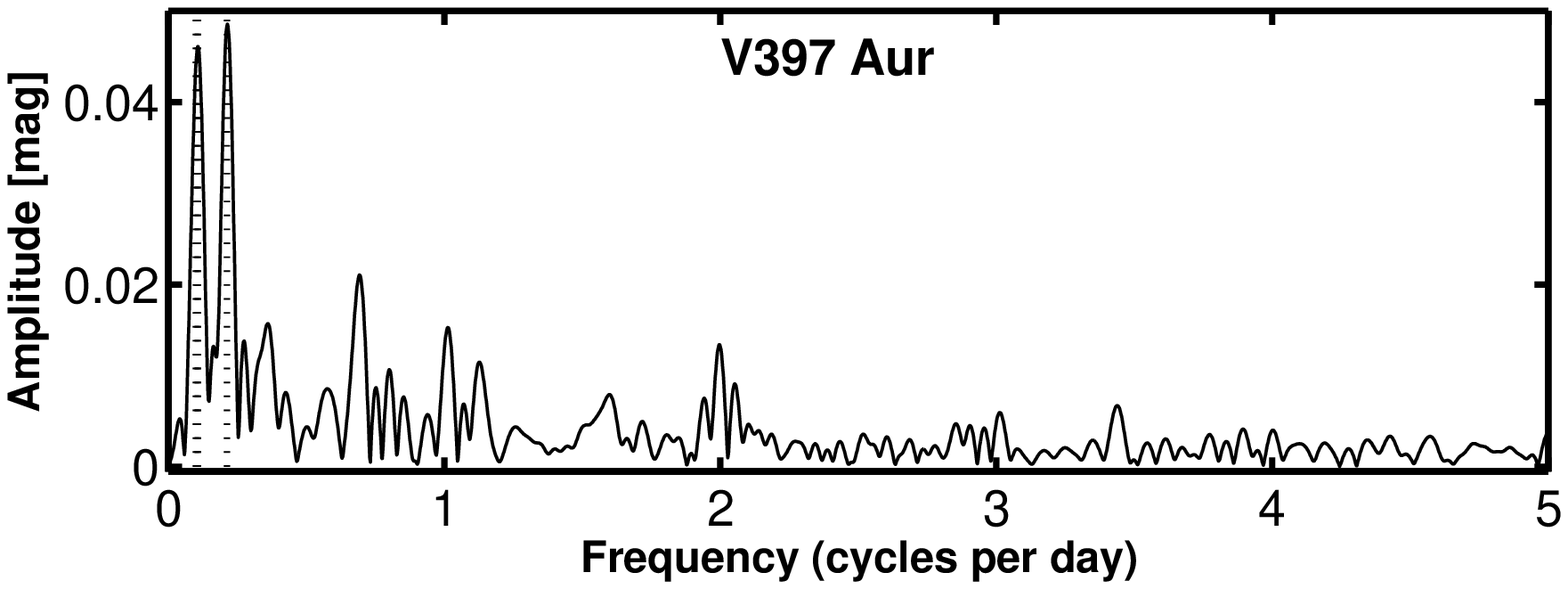}
\end{center}
\caption[]{\label{periodos} Periodograms, covering the 0--5 d$^{-1}$ frequency range. Vertical
dashed lines mark rotation rates reported in the literature; in the case of V396 and V397~Aur, these
closely match the position of our periodogram peaks.} 
\end{figure}

We used Period04 \citep{2005CoAst.146...53L} to produce periodograms for each light curve. This program 
handles data with non-uniform time sampling and performs a multiperiodic least-squares fit for frequency, 
amplitude, and phase of a series of peaks selected in the periodogram. We used Period04 in an iterative 
fashion: first searching for a single, most significant peak, then subtracting the corresponding best-fit 
sinusoid from the data, recomputing the periodogram, and searching for further peaks. We evaluated the 
significance of the identified signals by requiring that their amplitude be at least a factor of 4.0 
higher than the surrounding noise level in the periodogram for 99.9\% confidence, as suggested by 
\citet{1993A&A...271..482B}. In addition, a Monte Carlo simulation tool was used to add Gaussian noise to the 
best-fit combination of sinusoids, according to the noise level in the original light curve, and at the same time 
sampling. The periodogram was then regenerated, and the parameters of the highest peaks recorded.
We performed 500 realizations of this process to determine the uncertainties on the periods and 
amplitudes quoted in Table~2. The full set of periodograms is displayed in Fig.\ \ref{periodos}.

%\subsubsection{Rotation periods}
%\label{rotperiods}

Using Period04, we discovered a single statistically significant period for each star \citep[or two in the case of 
the V397~Aur system, which is a spectroscopic binary that was recently 
resolved;][]{1989AJ.....98..987M,2011ApJ...731....8K}, each of which is listed in Table~2. For V396~Aur, V397~Aur, 
HD~31305, the periodic pattern is relatively consistent over the 24-day duration of observations, and the light 
curves encompass up to 10 cycles. We therefore ascribe the variability to stable hot or cool spots on these objects 
and tentatively associate the detected periodicities with their rotation rates. SU~Aur also maintained stable 
periodic behavior consistent with spots, but only for the first $\sim$18 days before a dramatic fading event set it. 
The case of AB~Aur is even more complex, since its light curve displays systematic trends on top of the suspected 
periodicity. The light curve of this star is too erratic to conclude that its brightness fluctuations are reflective 
of its rotation period.

We list the values of all detected periodicities in Table 2. In several cases, they are consistent with 
one or more previously noted periods, suggesting that one or more spots are frequently present on the stellar 
surfaces and that the basic spot pattern is long-lived even if particular spots come and go over years. Nevertheless, 
the overall range of periods reported in the literature underlines how difficult it can be to infer accurate rotation 
rates from more sparsely sampled ground-based data.

To assess even shorter timescales of variability, we examined the full independent range of periodogram frequencies 
from 0 to 1000 d$^{-1}$ (or periods down to 1.5 minutes). We performed this analysis on residual data after the 
sinusoidal signals detected with Period04 were subtracted out from the light curves in order to exclude power from 
the dominant periodicities. In addition, we masked out the {\em MOST} orbital aliasing features at multiples of 
14.2~d$^{-1}$.

To model the frequency domain behavior, we fit a median trend to each periodogram and took its level at 
$f\sim$800--1000~d$^{-1}$ to represent the underlying white noise. These values ranged from 1.6$\times 
10^{-5}$ mag (HD~31305) to 3.5$\times 10^{-4}$ mag (V397~Aur) and, for our dataset, are 2--5\% 
of the photometric white noise in the time domain (i.e., the RMS light curve values listed in Table~1). We 
then fit a ``$1/f$'' curve (i.e., power proportional to inverse frequency, or amplitude proportional to $f^{-1/2}$) 
to the low-frequency end of the periodogram from 2--10~d$^{-1}$ and noted the point at 
which this exponential drop-off in variability amplitude with frequency reached the white noise limit. For 
AB~Aur, SU~Aur, and HD~31305, our targets with mmag-precision photometry, this occurred at $\sim$175--350~d$^{-1}$. 
We detect no substantial deviations from flickering behavior (apart from the aliasing power excess) 
on timescales down to several minutes. In addition, we can rule out high-frequency periodic signals with 
periods down to 1 minute and amplitudes down to four times the periodogram noise limit: 0.4~mmag for AB and 
SU~Aur, and 0.06 mmag for HD~31305. The results of our Fourier analysis suggest
that in general, observations detect flicker noise in CTTS down to the
white noise limit. Considering that ground-based photometry achieves a typical precision
5--10 times worse than our {\em MOST} uncertainties, future observations
would need to sample light curves roughly once per hour to capture the
essence of variability in these stars. 

For the two lower precision datasets on V396 and V397~Aur, amplitude levels in the periodogram followed a much shallower trend 
($\sim f^{-0.15}$) after removal of the main periodicities. We suspect that the light curves of these stars are fully explained 
by one or more spots along with low-level noise that is systematic but not characterized by flickering. We also fail to detect 
any significant high-frequency signals with amplitudes down to 1~mmag in any of our targets.  We concur from analysis of this 
sample with \citet{2010A&A...518A..54G}, who found no evidence for short-timescale periodicities due to oscillations of an 
accretion shock.

In summary, after removing the dominant periods due to rotation and repeating the Fourier analysis, the resulting 
periodograms of the CTTS were all relatively featureless and consistent with: a ``$1/f$'' flicker noise profile at 
low frequencies ($<$10~d$^{-1}$), a white noise baseline at high frequencies, and a slight power excess due to 
aliasing at intermediate frequencies. Thus it appears that there are no characteristic timescales underlying the 
variability in these objects, other than the rotation period. A representative example of the residual periodogram 
for AB~Aur is shown in Fig.\ 5.

\renewcommand{\thefootnote}{\arabic{footnote}}

\subsection{Wavelet analysis}

While each of our target stars displayed a prominent signal in the periodogram, we found that most of the lightcurves are 
better described as a mixture of periodic and aperiodic phenomena. To determine whether some of the stochastic behavior could 
be attributed to the appearance of coherent yet transient periodicities, we performed a wavelet transform using the WinWWZ 
package available through the American Association of Variable Star Observers\footnote{WinWWZ was produced by G. Klingenberg 
and L. Henkel; it can be downloaded from {\em http://www.aavso.org/winwwz}}. This program applies a weighted wavelet 
Z-transform \citep{1996AJ....112.1709F} to achieve resolution in both frequency and time for unevenly spaced data. The wavelet 
employed is a simple sinusoid plus constant term, applied by sliding a window of predetermined width across the data. 
Datapoints close to the center of the window have the highest weight, whereas those near the edges are downweighted. WinWWZ is 
a useful tool for not only understanding how many periodic signals are present in the data at a given time, but also how their 
frequencies and amplitudes evolve.

\begin{figure}[!h]
\begin{center}
\hspace{-1.1cm}
\includegraphics[scale=0.48]{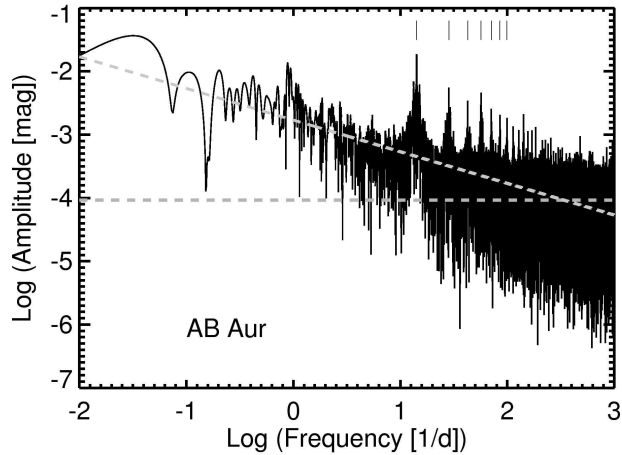}
\end{center}
\caption[]{Logarithmic periodogram of AB Aur, after a 6.5 day period is removed from the light curve.
We have fitted a ``1/$f$`` profile (i.e., amplitude proportional to $f^{-1/2}$) to the low-frequency regime (dashed diagonal line), 
and a uniform noise level at high frequencies (dashed constant line). The series of peaks starting at $\log$(frequency)=1.15 are 
aliases due to the {\em MOST} satellite orbital period; the first seven are marked with vertical ticks.}
\label{}
\end{figure}

The basic output of the wavelet transform consists of signal power as a function of time and frequency. We illustrate 
this for our target stars with a series of contour plots, shown in Fig.\ 6. WinWWZ also provides one-dimensional plots 
of frequency versus time for the strongest signal.  We detected one dominant period for each star (or two, in the case 
of binary system V397~Aur), affirming the results of Fourier analysis in Section 3.1. The derived values are presented in 
Table~2, with ranges indicating the minimum and maximum period values attained for this strongest signal over the course 
of the time series.

The wavelet analysis suggests that there may be changes in period over the course of the {\em MOST} 
observations. The abrupt disappearance of the periodic signal in the light curve and wavelet transform of 
SU~Aur around day 18 is the most obvious instance of a change in periodic behavior. Since we do not know a 
priori the shape of the subsequent brightness dip, it is difficult to ascertain whether the periodicity 
vanishes completely or is simply washed out by the strongly systematic light curve trend. If the former, then 
the disappearance of the signal indicates that the stellar photosphere becomes obscured by material in the 
line of sight. Other more subtle variations in period include a 1.5 day evolution in the strongest period of 
AB Aur, and a change of over 0.2 days in the period in the HD 31305. These effects may reflect actual changes 
in the variability timescale, or instead be attributed to non-sinusoidal components in the variability.

As the target with the most stable frequency behavior, V397~Aur shows two strong and sustained periodic 
signals in the wavelet analysis, consistent with its identification as a binary system.
For the other objects, the ranges in period measured for the dominant signal are of order 100 times larger than 
the 1-$\sigma$ uncertainties returned by the periodograms. Since the periodogram represents the average 
behavior of periodicities over the 24-day run, the wavelet result again suggests that there is subtle evolution in 
the timescales of periodic variability during our observations, or that aperiodic variability is interfering with
out ability to measure a stable period. 

\begin{figure*}[!h]
\begin{center}
\includegraphics[scale=0.65]{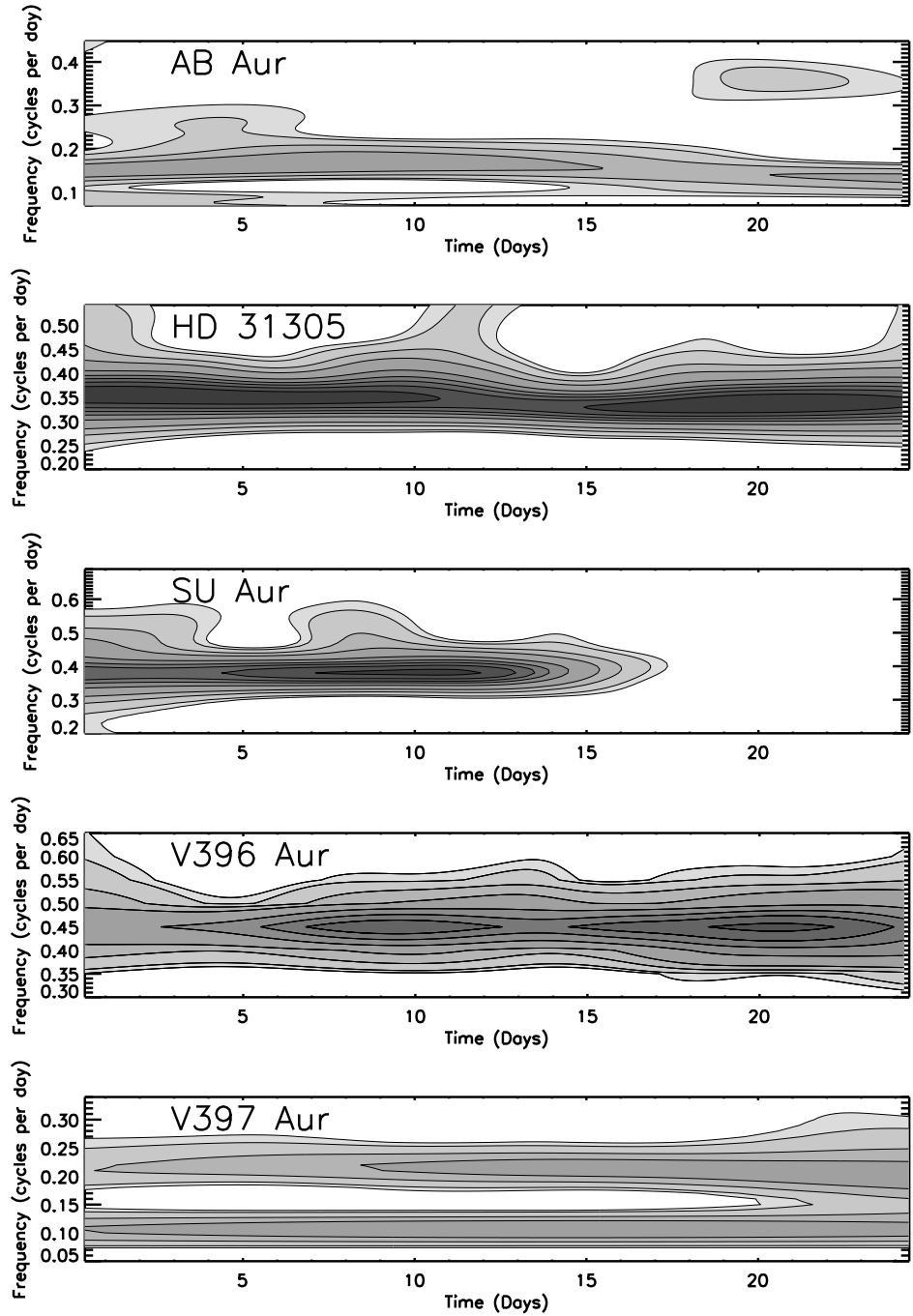}
\end{center}
\caption[]{Wavelets for all targets, zoomed in on the frequencies of significant signals. The intensity scaling 
is arbitrary but the same for all objects. The Fourier periods are recovered, but some additional structure is also apparent.}
\label{}
\end{figure*}

\subsection{Autocorrelation analysis}

Fourier analysis and wavelet analysis provided an indication of the periodic variations present in the data.  
Autocorrelation \citep{1976tsaf.conf.....B} is a further technique that enables confirmation of these results as 
well as additional searches for variability that is not strictly periodic, not sinusoidally shaped, or not 
present for the entirety of the data set.  It provides an assessment of how consistent variability patterns 
are on different timescales.  Perfect correlation results in an autocorrelation value of 1, while completely 
uncorrelated data returns a value of 0, and anticorrelated data corresponds to negative values.  This 
technique is particularly useful in that it can reveal periodic variations even if they are not persistent 
throughout the entirety of the data set.  It serves as a check on the results obtained by Fourier analysis 
and wavelet analysis.  Since the MOST data were not evenly spaced throughout the observation period, we first 
resampled the light curves using linear interpolation. We then applied the unbiased autocorrelation formula: 
$$A(t)=\frac{1}{\langle y^2\rangle j_{\rm max}}\sum_{j=1}^{j_{\rm max}}y(j)y(j+t/\Delta t),$$ where $N$ is 
the total number of data points, $\Delta t$ the timeshift, $y$ the
light curve values, and $j_{\rm 
max}=N-t/\Delta t$.

\begin{figure*}[!t]
\begin{center}
\epsscale{0.95}
\plottwo{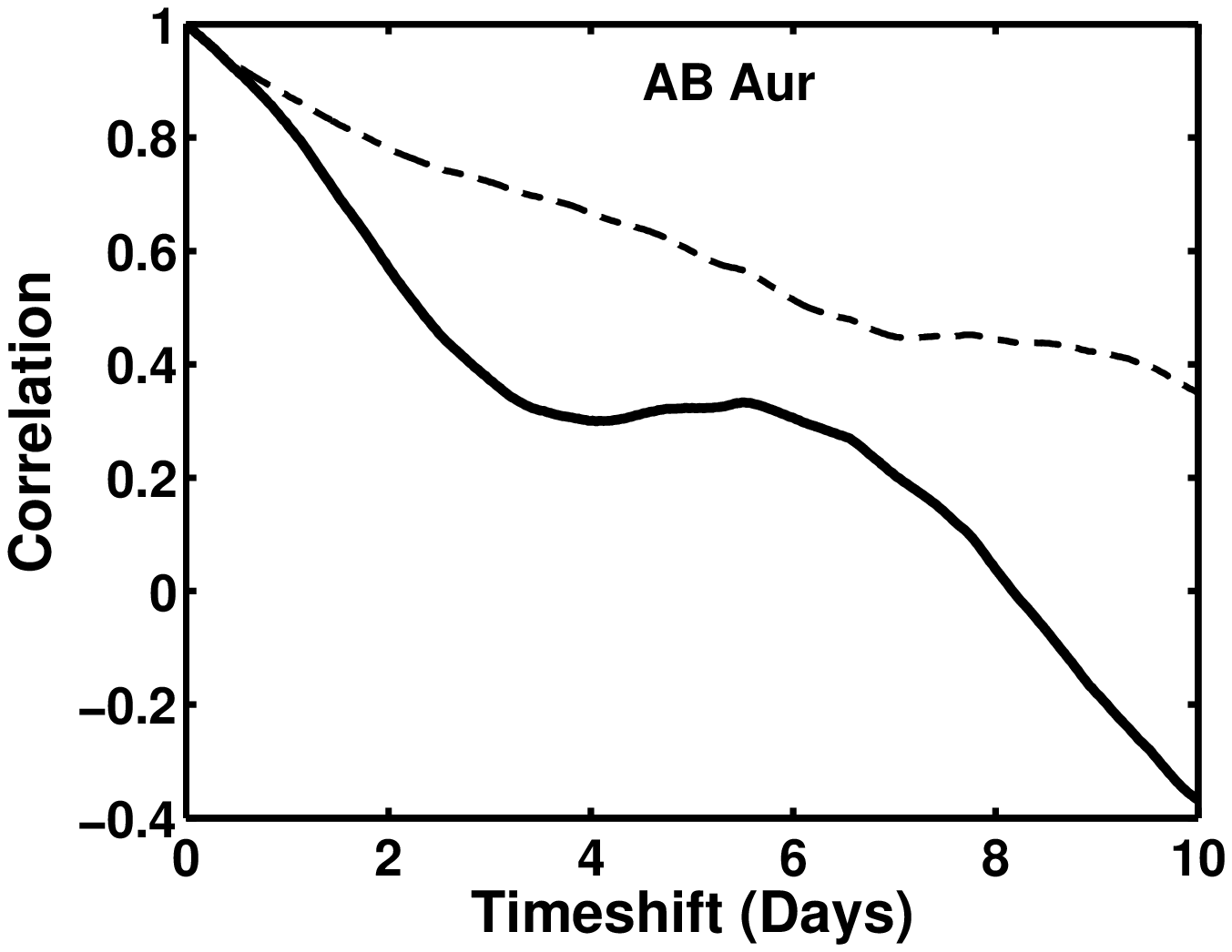}{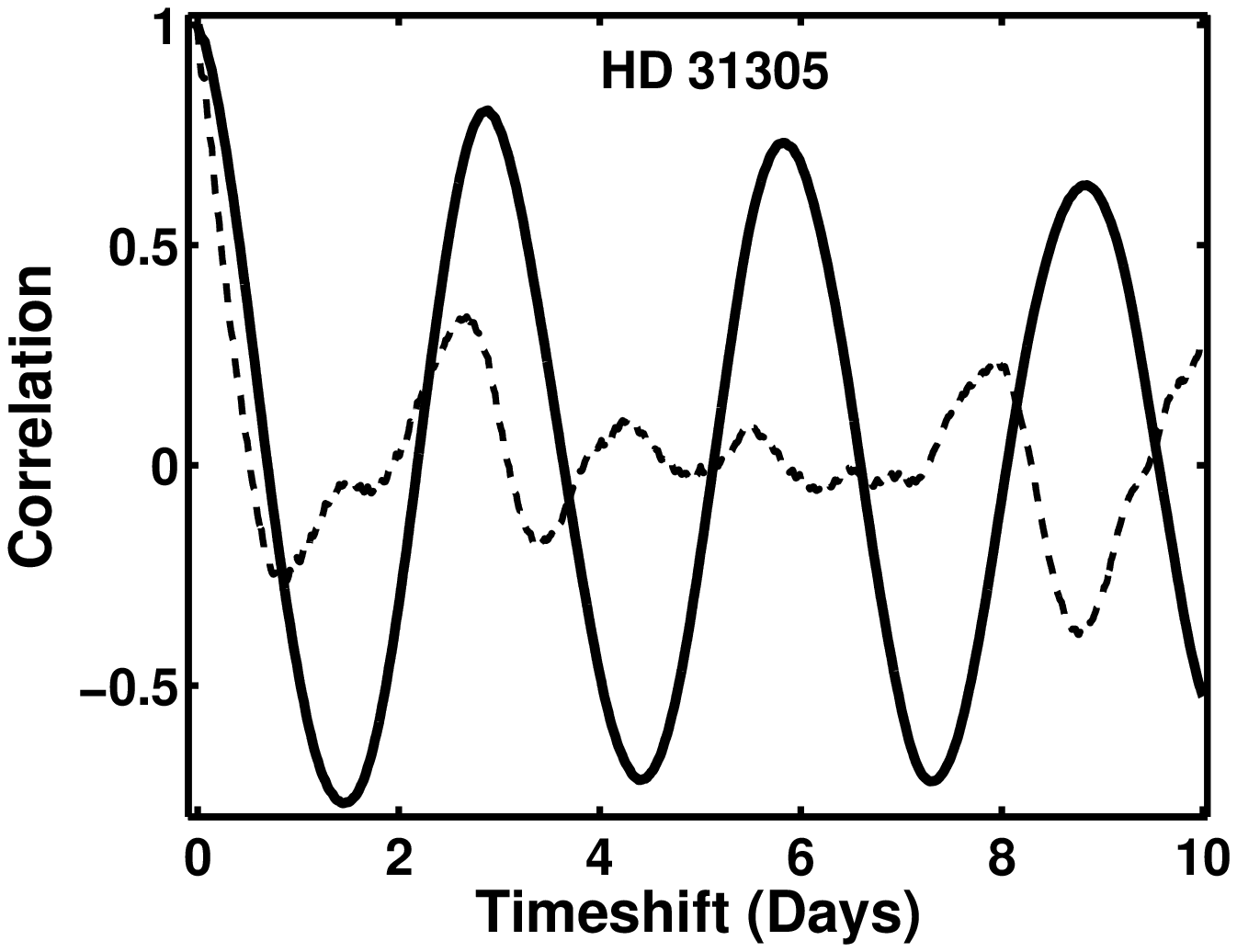}
\plottwo{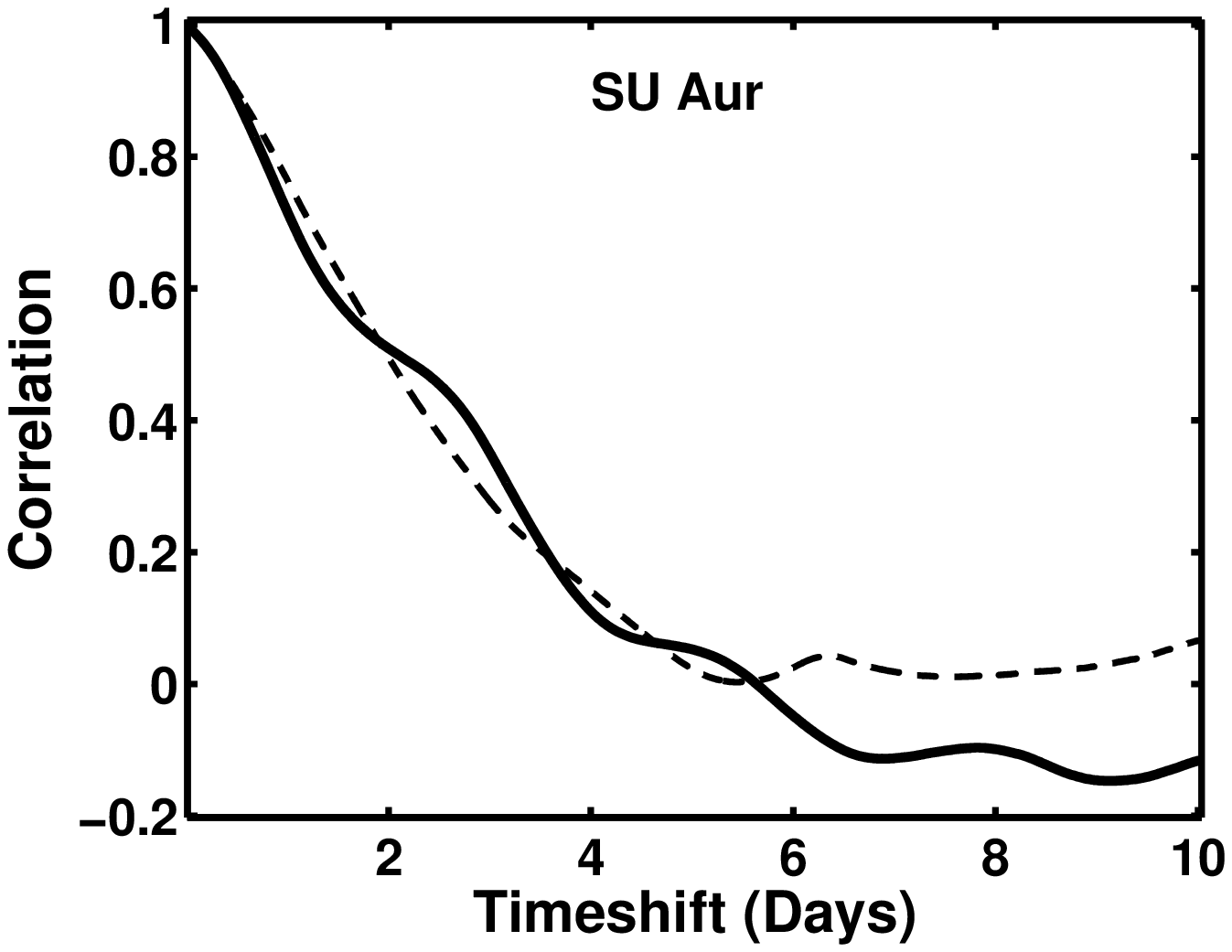}{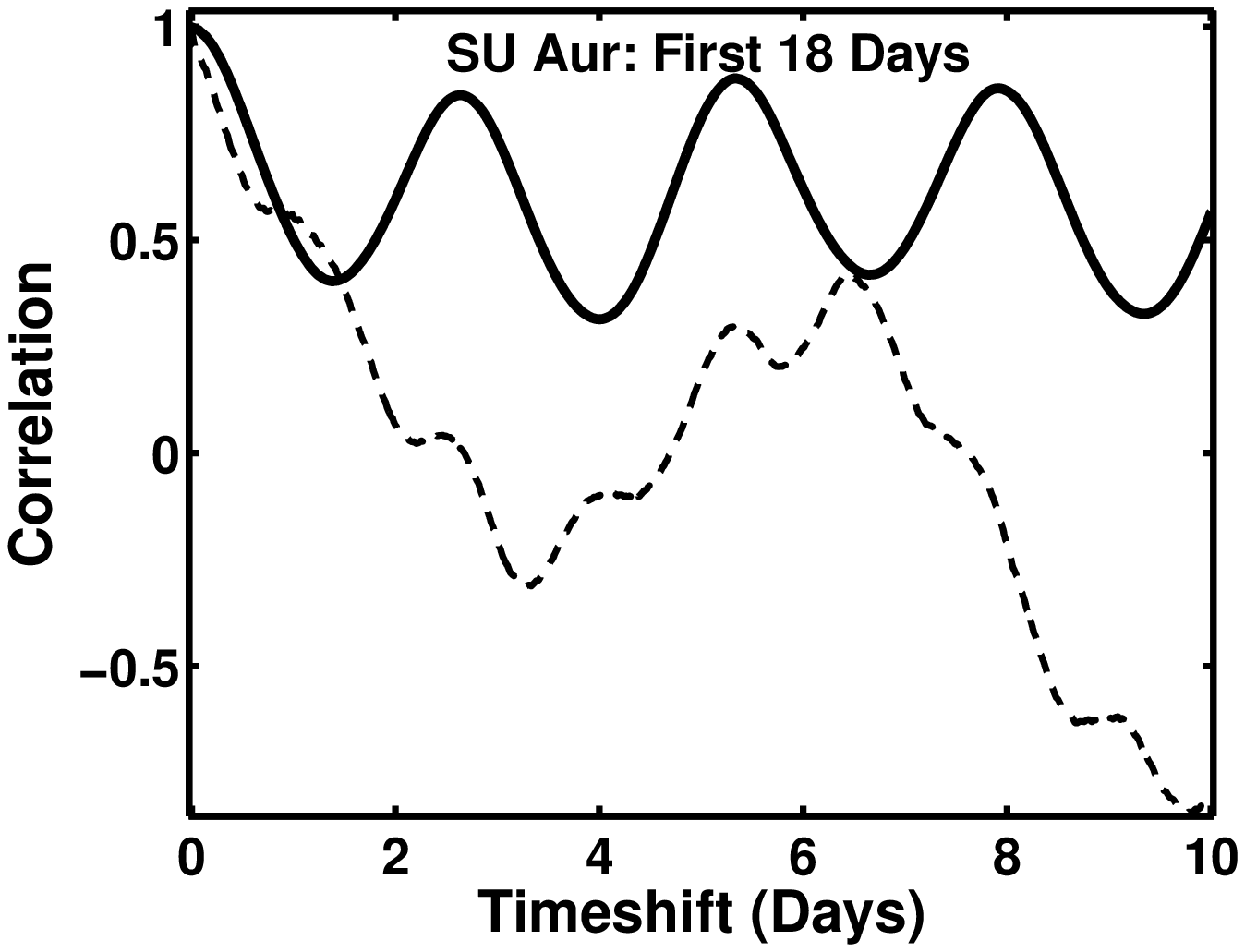}
\plottwo{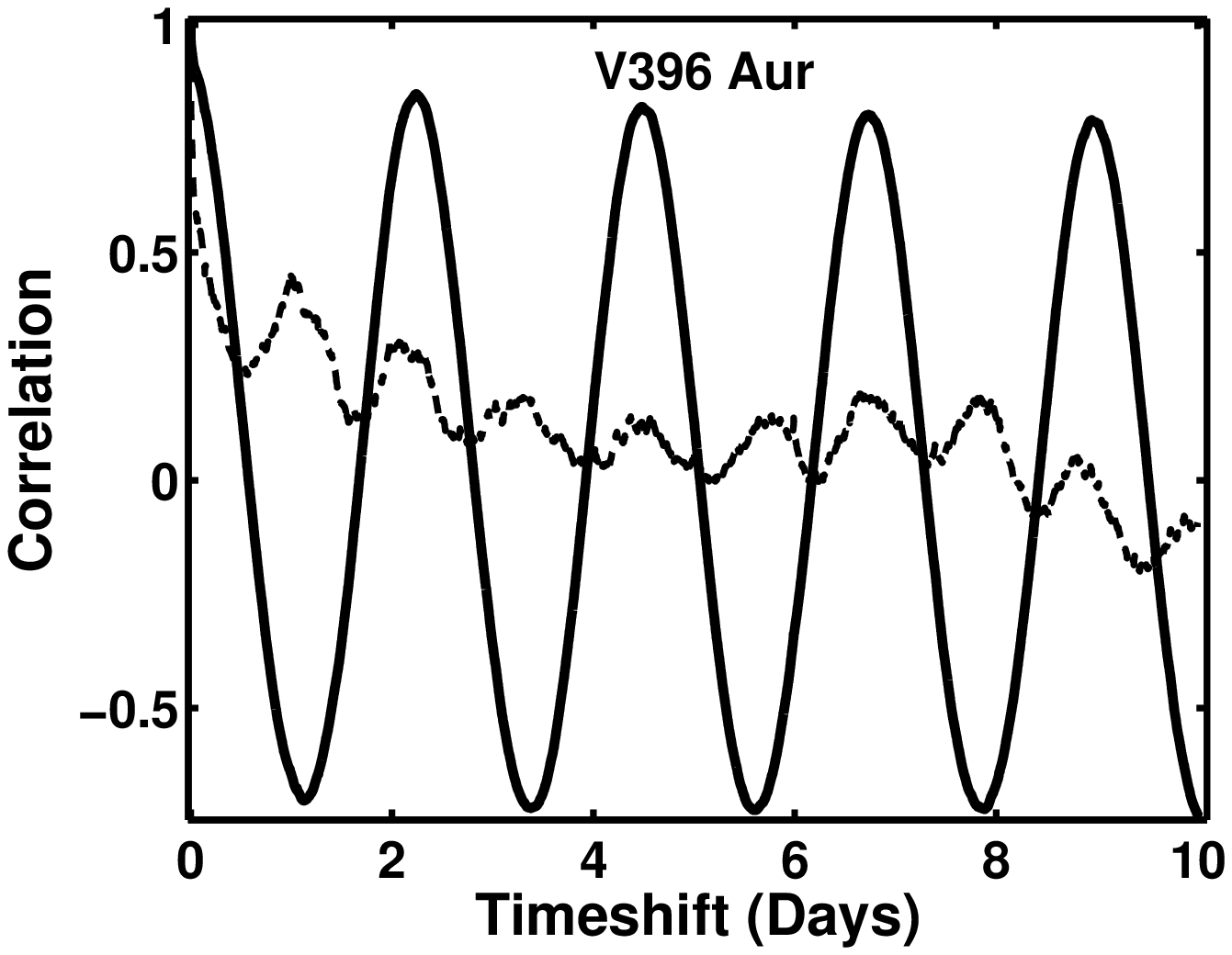}{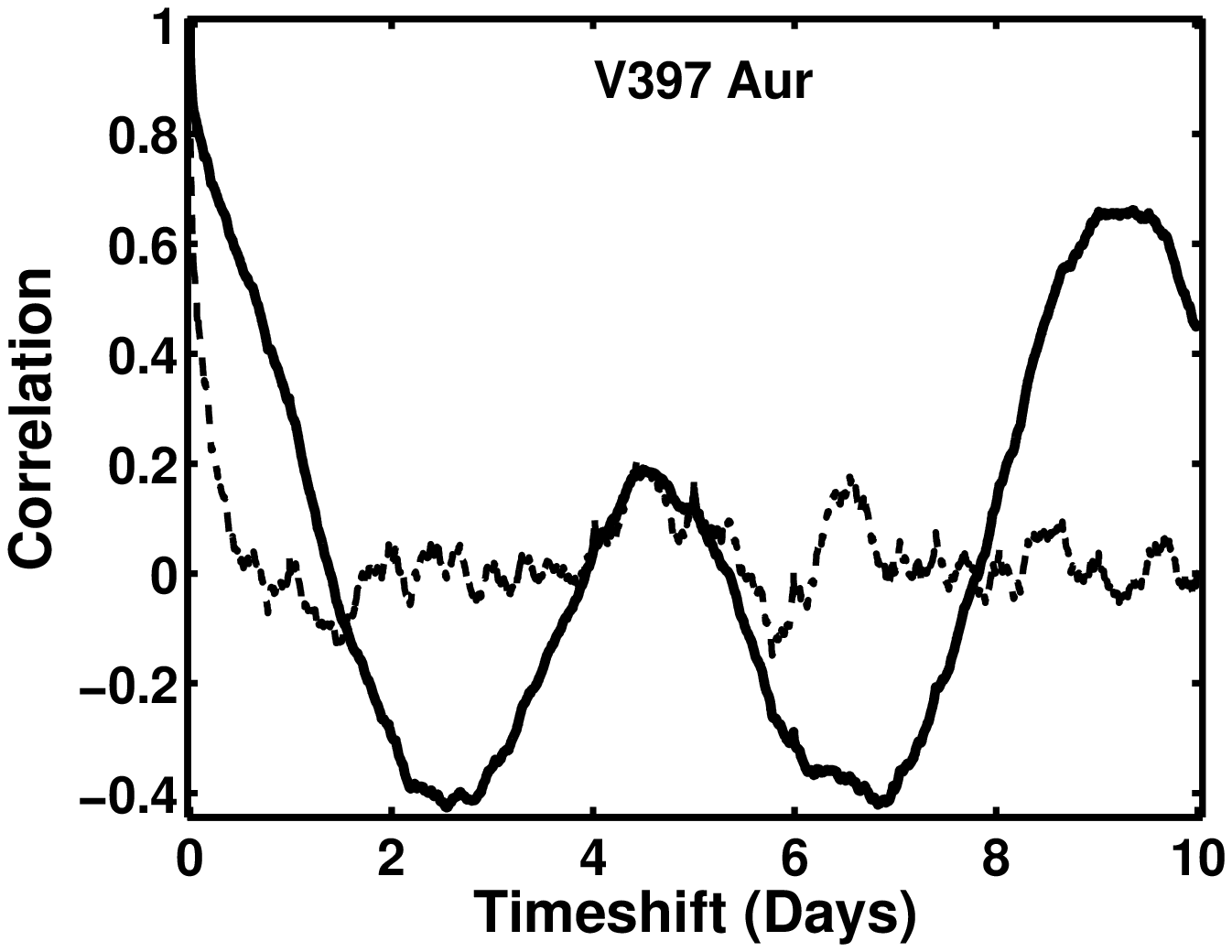}
\end{center}
\caption[]{Autocorrelations (solid curves) and residuals (dashed) after period removal. We have used the residuals to 
estimate a characteristic timescale for variability, by noting where the autocorrelations drop to 0.5. The second panel 
for SU~Aur is the autocorrelation performed on only the first 18 days of the time series, before the prominent fading 
event begins.}
\label{}
\end{figure*}

\begin{figure*}[!t]
\begin{center}
\plottwo{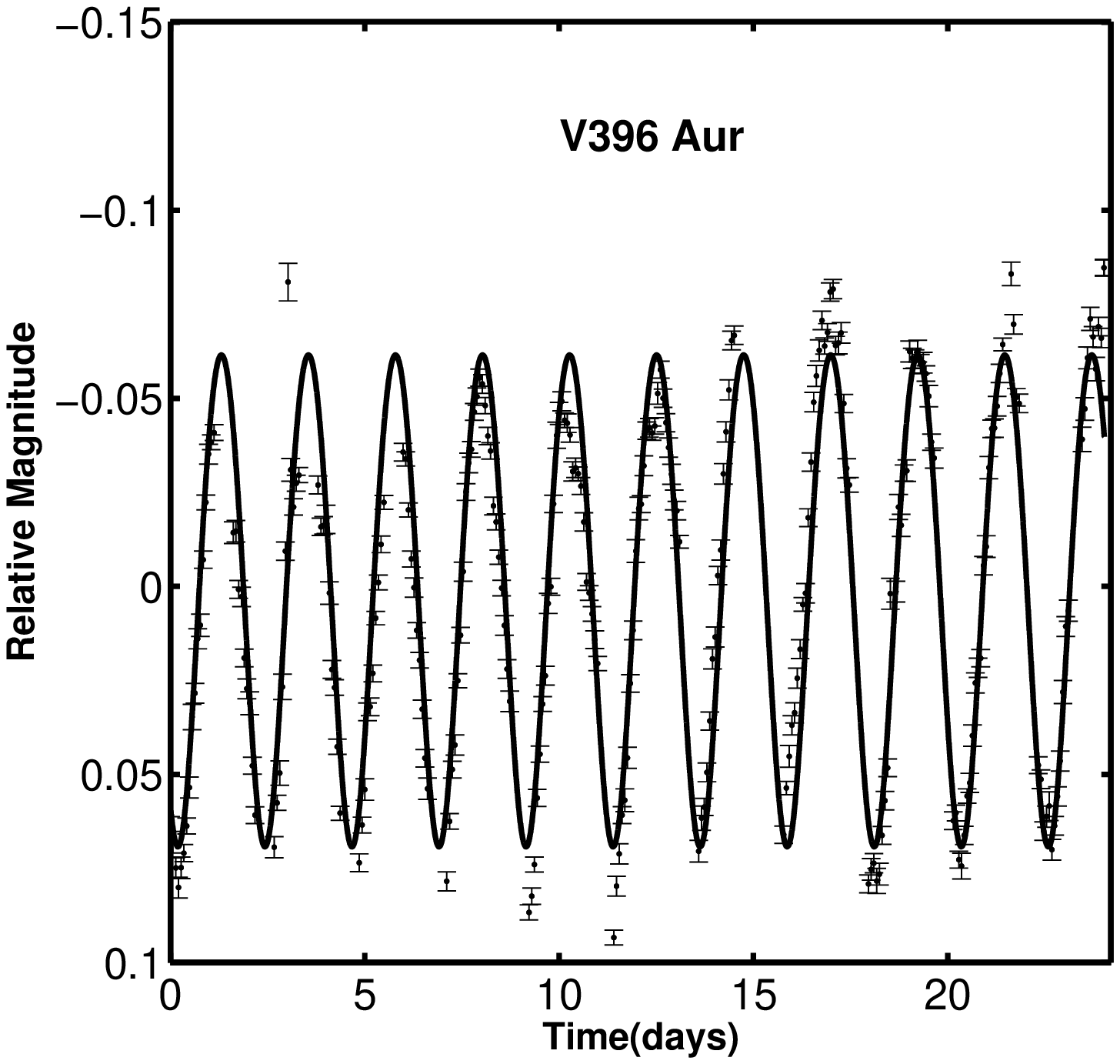}{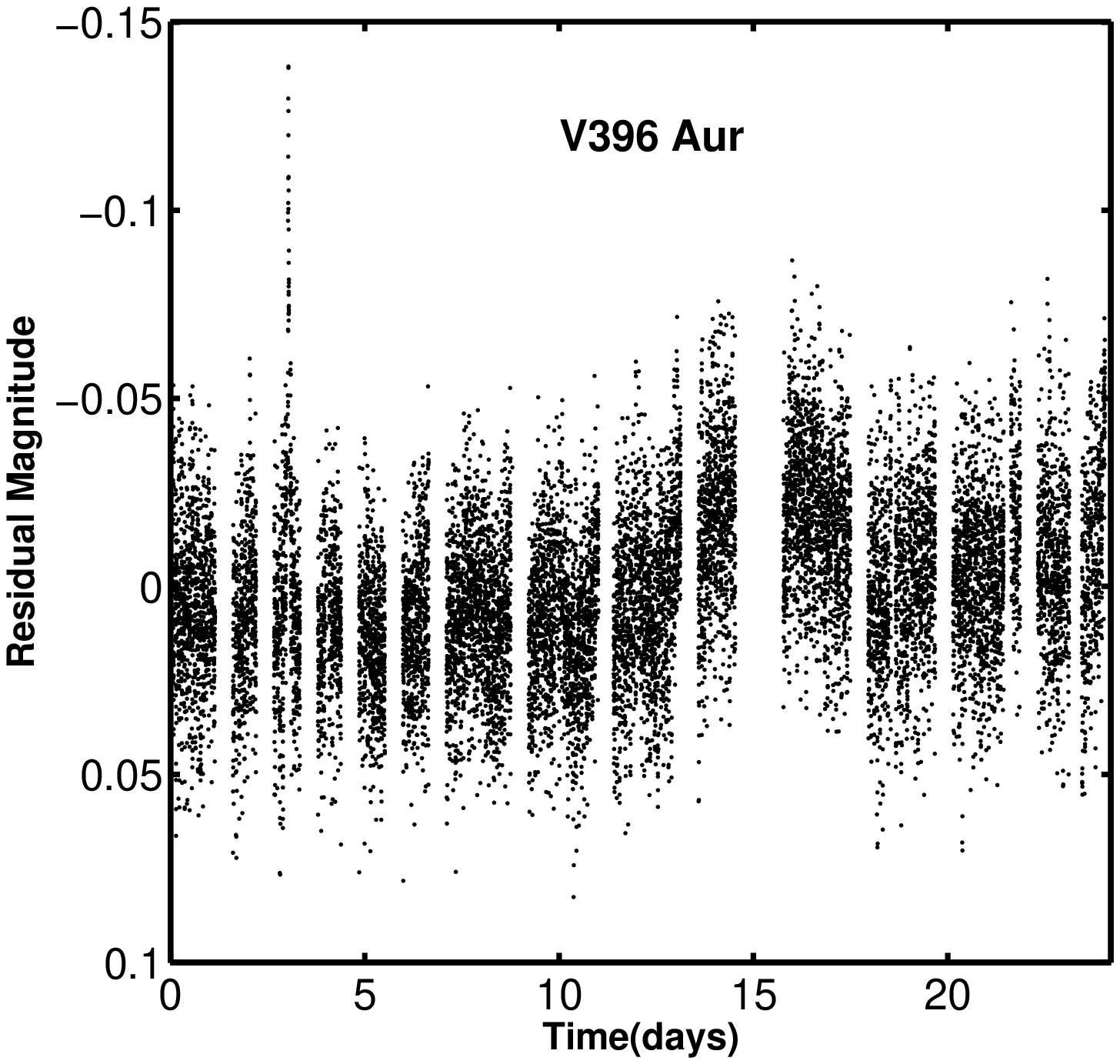}
\end{center}
\caption[]{\label{V396orbitbin} Left: Lightcurve of V396~Aur, binned on the {\em MOST} orbital timescale (points
with uncertainties), compared to one of the best fitting spot models (solid curve). Right: Unbinned light curve
residuals after model subtraction.}
\end{figure*}

We noted a single prominent autocorrelation peak for each star, consistent with the periods derived from Fourier 
analysis. We have derived periods by measuring the timeshift of the first maximum and noting the surrounding values 
for which the autocorrelation function drops by 1\%; these ranges are listed in Table~2. Since the light curve of 
SU~Aur ceases to display oscillations during the last $\sim$6 days of observation, we have computed two versions of 
the autocorrelation function for this star. The first is performed on the entire dataset, while the second includes 
only the first 18 days.

In addition to providing confirmation of the previously measured periods, autocorrelation analysis enables 
investigation of any further underlying variability. To accomplish this, we first used Period04 to subtract 
out variability at the dominant period. We then recomputed the autocorrelation function on this residual. 
Both versions of the autocorrelation are shown in Fig.\ 7. The residuals contain low-level undulating features, but 
further subtraction of periodogram-detected signals from the light curve did not result in their disappearance.

While no significant new variability signals were evident in the residual autocorrelations, the analysis does provide 
an indication of the characteristic timescales for aperiodic physical processes in our target stars.  We associate an 
autocorrelation timescale by noting the value for which the autocorrelation drops to 0.5. The flux variations of HD 
31305 are correlated on timescales of $\sim$0.2 day once the periodic signals are removed, while SU Aurigae is 
correlated on timescales of $\sim$1.1 days. In contrast, AB Aurigae displays coherence in its variations out to 6.2 
days. Long timescale correlation after the periodic signal has been removed implies that for these three sources there 
are additional physical processes other than periodic spot modulation that contribute to the variability. The residual 
light curves of the two weak-lined T Tauri stars V396 and V397~Aur are correlated on much shorter timescales of 
$\lesssim$0.1 days. We suspect that spots explain most of the variability in these systems.

\section{Starspot modeling}

The periodogram, wavelet, and autocorrelation analysis methods yielded one significant periodicity per star.  
In most cases, the derived timescale was very close to a rotation period previously reported in the literature.  
The long-term stability of these periodicities, as well as the deviations in light curve shape on shorter time 
scales, suggests that the underlying mechanism is flux modulation by cool or hot starspots, as opposed to 
stellar pulsation or an eclipsing companion.

Representation of a light curve with spots requires a number of parameters and often admits degenerate 
solutions. A non-parametric model involving only total spot coverage and temperature contrast relative to 
the photosphere is one way to avoid these challenges \citep[e.g.,][]{2008A&A...479..827G}. However, the dense 
sampling and long duration of our time series observations enables more detailed exploration of the presumed 
spot properties than is typically possible with ground-based data. We exploit the high quality of our dataset 
by applying StarSpotz \citep{2006AAS...20923002C,2007ApJ...659.1611W}, an analytic starspot modeling program 
designed specifically for the analysis of {\em MOST} data.

To reduce the number of free parameters required for the spot modeling process, we fixed the stellar inclination, 
$i$, as well as the rotation period. We set the latter according to our results from Fourier analysis (Section 3.1). The 
value of $i$ was taken from the literature where available. In addition, we fixed the limb darkening coefficient via 
the $V$-band prescription of \citet{2011A&A...529A..75C}, but tested several alternate values to determine the 
sensitivity of our results to this parameter.

We provided starting guesses for all other free parameters, including unspotted stellar brightness, spot darkness 
(ratio of flux compared to that of the surrounding photosphere), spot latitude and longitude, as well as spot 
angular size. To avoid unphysical values of spot darkness, we produced models for discrete steps between 0.0 
(completely black spot) and 1.0 (same brightness as photosphere). The total number of spots, as well as the presence 
of differential rotation, were specified in advance, and the free parameters were then varied separately for each 
spot to create a suite of analytical models following the prescriptions of \citet{1977Ap&SS..48..207B} and 
\citet{1987ApJ...320..756D}. Starspotz evaluates the goodness of model fits to the light curve via a non-linear 
least squares algorithm, returning a local minimum $\chi^2$ value.

We assessed the improvement in fit brought about by increasing the number of spots, as well as whether an assumption of 
differential rotation produced a better fit. In general, neither of these additions resulted in significant improvement 
and there remained deviations in the light curves that could not be explained by spots. Since these additional features 
are lower in amplitude than the periodic variability, we argue that spot models can help nevertheless to inform our 
picture of stellar surface properties. In the following sections, we detail the the best-fit spot configurations and 
evaluate their consistency with the light curves of V396~Aur, SU~Aur, HD~31305, and AB~Aur. We did not produce a spot 
model for V397~Aur since it is a binary system and both members appear to contribute to flux variations in the light curve.

\subsection{V396 Aurigae}

Several rotation periods for the star V396~Aur were previously reported in the literature.  The 2.228 days by 
\citet{2008A&A...479..827G} and 2.24 days by \citet{1993A&A...272..176B} agree well with our values derived from 
Fourier, wavelet, and autocorrelation analysis.  The former study, known as the Maidanak survey, characterized 
variability in this object as having a correlation between the magnitudes of maximum and minimum light during 
different observing seasons. An explanation for this behavior could be long term evolution of the total spot area, 
which they represent by a collection of cool spots on the stellar surface. Despite the history of observation, the 
typical distribution of the starspots causing variability in V396~Aur remains ill determined. In contrast to 
\citet{2008A&A...479..827G}'s conclusions, \citet{2008ApJ...678..472H} proposed a single large spot near the star's 
pole, based on 2.18-day periodic radial velocity variations of $\sim$1 km~s$^{-1}$ amplitude.

We set out to model the starspot configuration of V396~Aur during our {\em MOST} observations by fixing known 
stellar parameters and generating a series of one to three dark spots. The $v$sin$i$ value of V396~Aur is 
18.6$\pm$1.9 km~s$^{-1}$ \citep{1987AJ.....93..907H}, yielding stellar inclinations from $\sim$25--35\arcdeg\ for 
the range of range of radii expected for a young K0 star \citep[$R\sim$1.5~$R_\odot$;][]{1995A&A...299...89B}. To 
generate Starspotz models, we adopted the two most extreme values reported in the literature: $i$=33\arcdeg\ 
\citep{1995A&A...299...89B} and $i$=25\arcdeg\ \citep{2008ApJ...678..472H}, along with a limb darkening coefficient 
$u=0.66$ appropriate for the K0 spectral type of V396~Aur ($T_{\rm eff}\sim$5250~K) from 
\citet{2011A&A...529A..75C}. Combined with the adopted $v$sin$i$ and stellar radius, our observed photometric period 
of 2.24~d is consistent with values of $i$ in the range 33\arcdeg$\pm$4.

We find that the {\em MOST} observations of V396~Aurigae are best modeled by a single large spot at 60--80\arcdeg\ 
latitude, with the lower latitudes favored for the stellar inclination $i$=25\arcdeg.
The best fitting solution is dependent on the spot darkness parameter, with darker spots corresponding to smaller 
covering fractions. We found a minimum spot diameter of 23\arcdeg, or $\sim$4\% of the stellar surface area). We find 
acceptable values for the spot darkness between 0.0 and 0.5; spots with lower contrast would occupy an unphysically 
large portion ($>$30\%) of the visible photosphere. The solution is also degenerate with the unspotted stellar 
brightness, again leading to very large spots for unspotted
brightnesses much greater than the observed flux level. Incorporation of two or more spots to the model did not result in a 
significantly better fit to the light curve. Adjustments to the limb darkening coefficient also failed to bring about 
improvement.

In Fig.\ \ref{V396orbitbin} we overplot one of the best fitting models on the light curve of V396~Aur, with data binned 
on the 101-minute {\em MOST} satellite orbital timescale, alongside the residuals of the unbinned light curve after 
model subtraction. There are a number of outliers from the model, one of which occurs at $t$=3 days into the run. Closer 
inspection of the full cadence light curve reveals that this is a flare event (see Fig.\ 2). The rise and decay times 
are 0.5 and 1.2 hours, respectively, and the amplitude is $\sim$0.1--0.15 mag. Other outliers lie mainly at the 
peaks and troughs of the light curve, suggesting that spot evolution is taking place, or that there is an additional 
source of low-level variability that is most apparent at these phases. Nevertheless, there do not appear to be any 
systematic trends or periodicities in the residuals. The minimum reduced $\chi^2$ value was 36.1, confirming that a spot 
model does not provide a full explanation for the variability in V396~Aur.

\subsection{SU Aurigae}

\begin{figure*}[!t]
\begin{center}
\plottwo{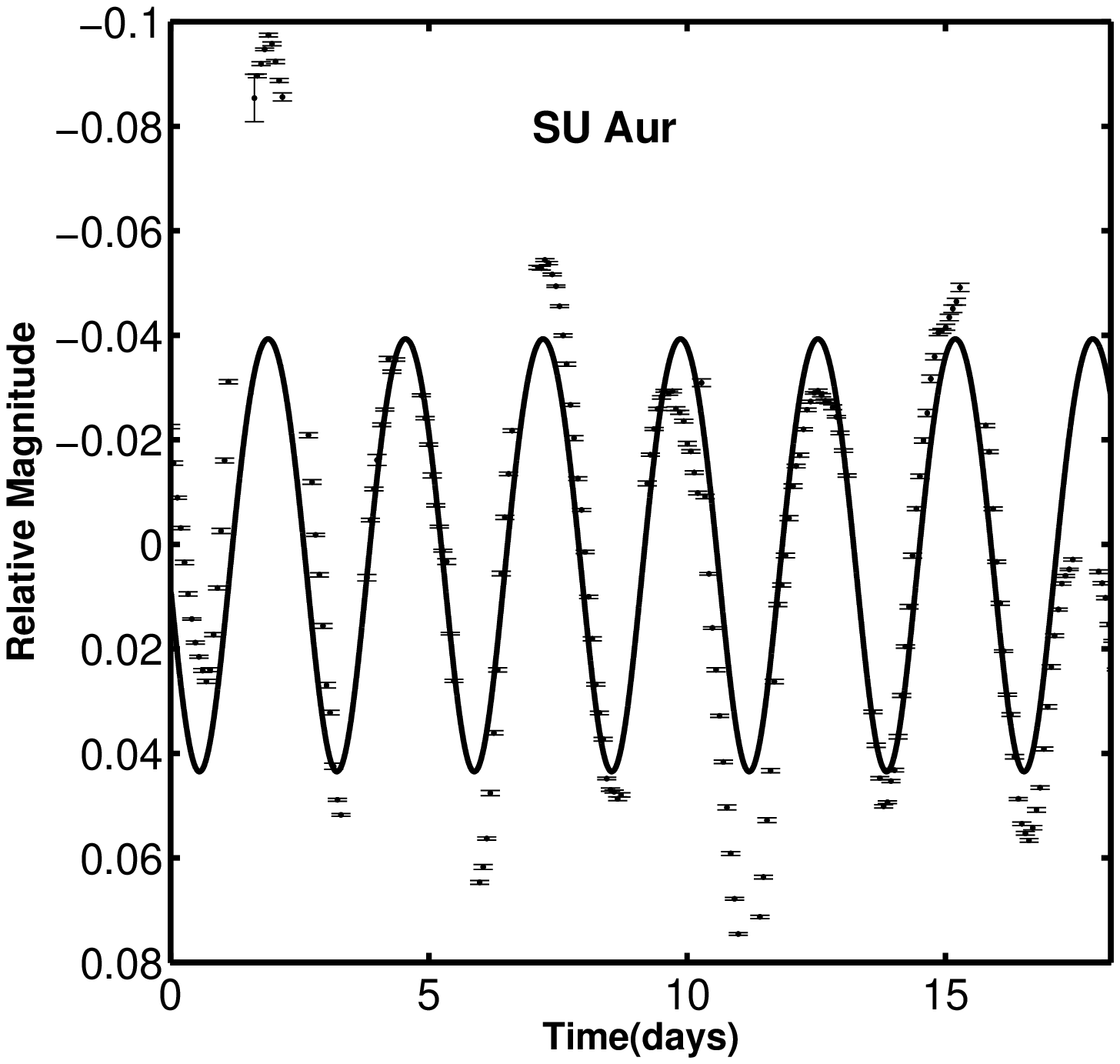}{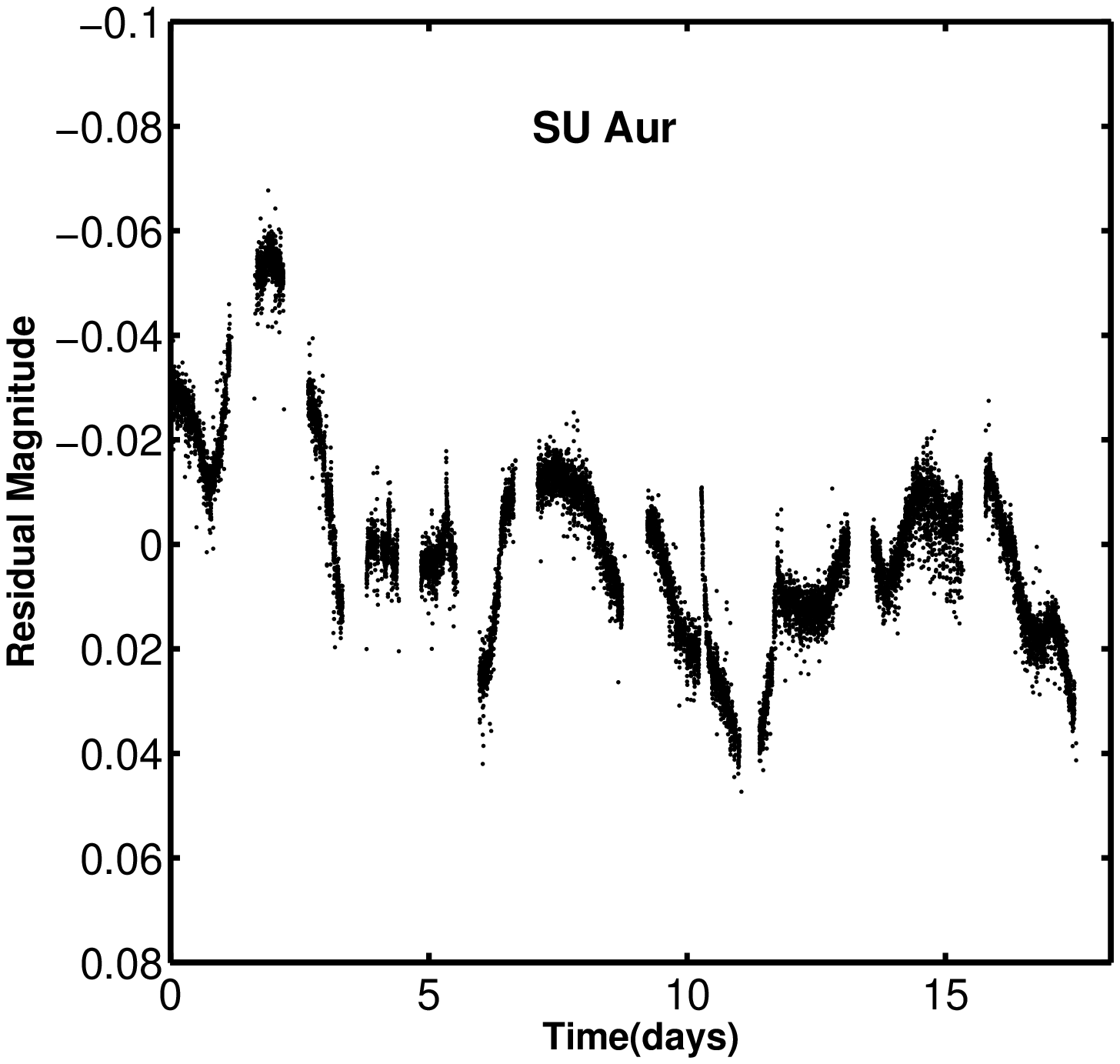}
\end{center}
\caption[]{Left: Lightcurve of SU~Aur, binned on the {\em MOST} orbital timescale (points with uncertainties), compared
to one of the best fitting spot models (solid curve). Right: Unbinned light curve residuals after model subtraction.}
\label{}
\end{figure*}

Previous observation of the G2 star SU Aurigae has indicated periodic variability on a variety of derived time 
scales (Table 1). A number of observers have also noted its erratic dimming by $\sim$0.5 mag in the 
optical over timeframes of 3--5 days \citep{2004MNRAS.348.1301U}. This has led to its classification as a 
UX~Orionis star (``UXOR''), although the change in magnitude is not as large as that of prototypical UXORs 
($\sim$1 mag in {\em V}; Herbst \& Schevchenko 1999) or slightly lower mass ``Type III'' T Tauri stars 
($\lesssim$1 mag; Herbst et al 1994).

%Mennessier 1997: Doppler imaging - any other sources?

Our observations of SU Aurigae also indicate a picture of variability involving a mixture of behavior. We observe both a 
periodic component during the first 20 days of observation and an 0.4 mag dimming event that begins around 
$t$=19~d and persists through the end of the time series.  In addition, a flare occurs at $t\sim 10.2$~d, with rise time 
$\sim$1 hour, decay time $\sim$2.4 hours, amplitude 0.03--0.04 mag (see Fig.\ 2). At $t\sim 1.7$~d, a portion of 
the light curve appears offset and brighter by $\sim$0.04 mag. As far as we can tell, this is not an instrumental 
error, and the star was inexplicably brighter during this particular peak of its oscillatory pattern.

To model the periodic variation in SU~Aur, we have considered only the first 20 days of the time series, since the last 
five days are dominated by systematic fading. Our chosen limb darkening coefficient is $u=0.63$ based on 
\citet{2011A&A...529A..75C}, and appropriate for a young G2 star \citep[$T_{\rm eff}\sim 
5550$~K;][]{2003ApJ...590..357D}. We chose a stellar inclination by assuming that it is equal to the inclination 
measured interferometrically for the disk: $i=62^{+4}_{-8}$\arcdeg\ \citep{2002ApJ...566.1124A}. The spectroscopic 
rotation velocity has been consistently measured at $\sim$66 km~s$^{-1}$ \citep{1986A&A...165..110B,1989AJ.....97..873H, 
2005ESASP.560..571G}. We adopt the value of $v$sin$i$=66.2$\pm$4.6 km~s$^{-1}$ reported by \citet{1986ApJ...309..275H}. 
Combining these measurements with values of the stellar radius reported in the literature ($R=2.75\pm 0.25 R_\odot$ from 
\citet{2003ApJ...590..357D} and $R\sim$3.6~$R_\odot$ from \citet{1996A&A...314..821P}), we find a range of expected 
rotation periods from 1.9--2.4 days. Including propagation of uncertainties in inclination and rotational velocity, this 
range expands to 1.7--2.6 days. The inferred periods are marginally inconsistent with the 2.66d timescale that we have 
measured photometrically.

Assuming that the variability is caused by features on the stellar surface, we find that the periodic component of the 
light curve could be accounted for by either a dark or bright spot at high latitude (80--85\arcdeg). For the dark spot, 
we determine plausible ratios of spot to photospheric flux between 0.0 and 0.3, and a spot size of $\sim$28--40\arcdeg\ 
($\sim$6--12\% of the total stellar surface area). A bright spot, on the other hand, could be a factor of 2.0 or more 
brighter than the surrounding photosphere and up to 30--40\arcdeg\ in size (7--12\%).

As with V396~Aur, neither the addition of more spots nor adjustments in the limb darkening parameter offers substantial 
improvements in the model fit. We note that with the single band {\em MOST} observations, we are unable to distinguish a 
hot spot generated by an accretion shock from a cool spot generated by the stellar magnetic field. However, the unstable 
amplitude of variability from one cycle to the next in our light curve suggests that accretion may be a better 
explanation. Since the model light curves appear nearly identical, we overplot only the dark spot model on the 
orbit-binned light curve in Fig.\ 9, alongside the residuals after model subtraction. The full light curve of SU~Aur 
shown in Fig.\ 1 displays inconsistent amplitudes from one cycle to the next. Examination of the residuals in Fig.\ 9 
reveals that there are both positive and negative deviations, with no systematic trend. The minimum reduced $\chi^2$ 
value is over 4000 for all models, indicating a poor fit. We therefore suspect that there is an additional source of 
variability present, or that the periodic mechanism involves dynamic evolution on $\sim$1 day timescales. In Section 5.3 we 
propose that orbiting material at the inner disk edge of SU~Aur offers an additional explanation for the observed 
periodicity.

\subsection{HD 31305}

\begin{figure*}[!t]
\begin{center}
\plottwo{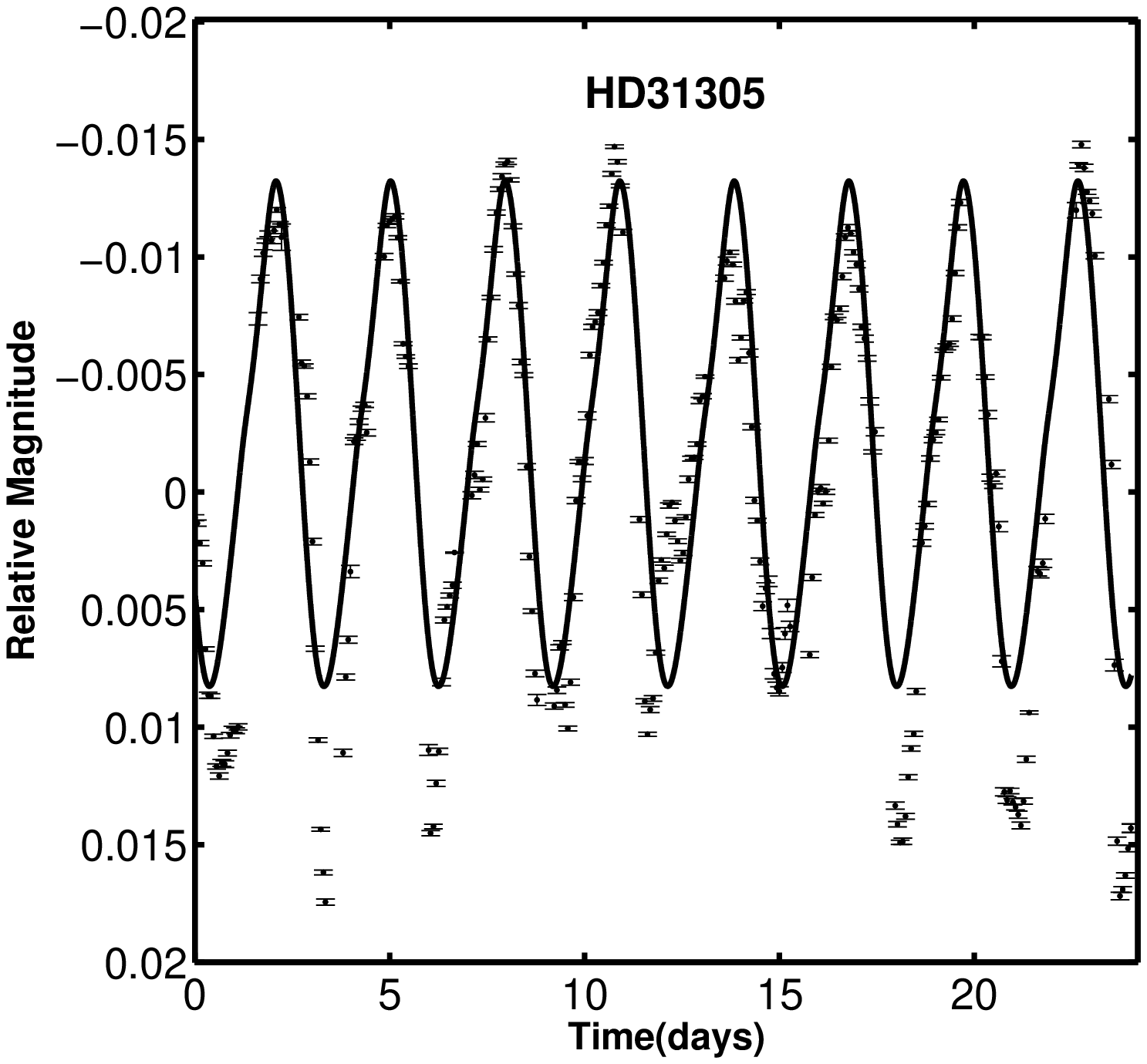}{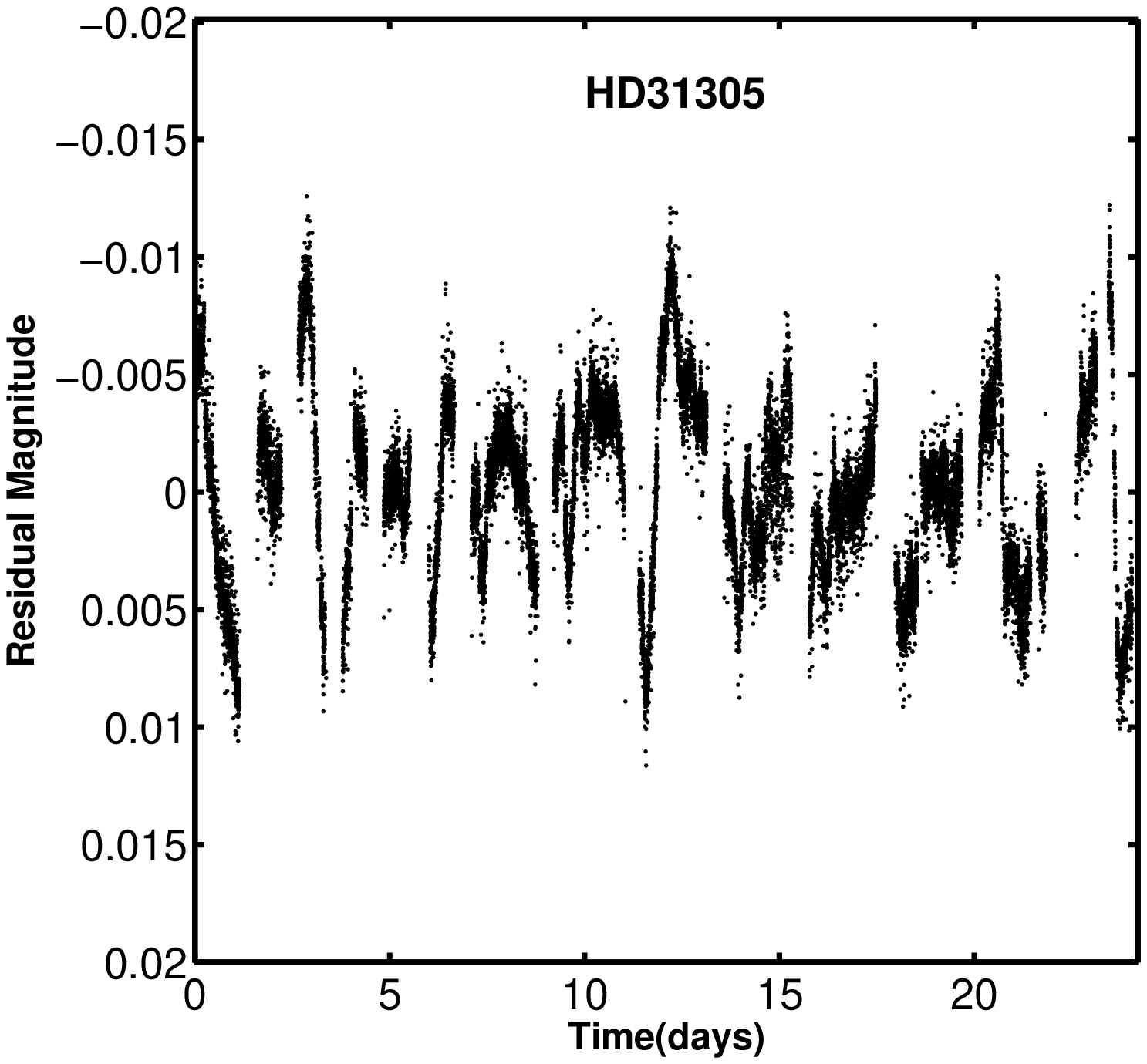}
\end{center}
\caption[]{Left: Lightcurve of HD~31305, binned on the {\em MOST} orbital timescale (points with uncertainties), compared
to one of the best fitting spot models (solid curve). Right: Unbinned light curve residuals after model subtraction.}
\label{}
\end{figure*}

HD~31305 is not a known member of the Taurus-Auriga star-forming region, but we include it in our presentation since it 
is also a bright star in our {\em MOST} field, and it exhibited surprising periodic variability. Full interpretation of 
the variability observed with {\em MOST} hinges on the status of this A0 target as a potential young star, a question 
which we defer to Section~5.2. A series of pronounced cyclic flux
changes suggest that either an inhomogeneous stellar 
surface or orbiting circumstellar disk features play a role in the variability. Stellar pulsation is another periodic 
phenomenon present in early-type stars, but obvious inter-cycle changes in the light curve do not support this 
alternative explanation. With a period near 2.8 days and amplitude of $\sim$0.01 mag, the variability pattern 
displays significant deviation from sinusoidal behavior at the 0.5\% level, none of which corresponds to significant 
periodicities in the periodogram.

We have attempted to model the overall behavior of the light curve with a series of dark or bright starspots. These 
could represent either regions of cooler surface temperature, patchy photospheric chemical composition such as observed 
in some peculiar type A stars \citep[e.g.,][]{2010A&A...524A..66S,2011IAUS..273..249K}, or hot spots related to 
low-level accretion. The inclination of HD~31305 is unknown, so we estimate it by considering the stellar parameters in 
concert with the $v$sin$i$ value of 50$\pm$25~km~s$^{-1}$ measured by Mooley et al.\ (2013, in preparation). Of note, this 
$v$sin$i$ puts HD~31305 at the low end of the distribution of rotation rates for A stars \citep[see][, Fig.\ 
7]{2012A&A...537A.120Z}. Using $U$, $B$, $V$, $J$, $H$, and $K$ magnitudes from the Simbad database along with a 
spectral type of A0, we derived a bolometric luminosity based on SED fitting. Using only the optical magnitudes, we 
derive a luminosity of log($L/L_\odot$)=1.38. Since there is an infrared excess associated with this object, 
non-photopheric emission likely contaminates the SED at $J$ band and beyond; incorporating the magnitudes here results 
in a larger luminosity of log($L/L_\odot$)=1.5. This range of luminosities, along with an A0 effective temperature of 
$\sim$9850~K results in a stellar radius of 1.85$\pm$0.09~$R_\odot$. If the observed periodic variability is due to 
surface starspots, then the combination of the 2.97~d period and estimated radius require a $v$sin$i$ of at most 
31~km~s$^{-1}$. For a minimum $v$sin$i$ of 20~km~s$^{-1}$, we derive $i$=39\arcdeg. To perform the spot modeling, we 
therefore adopted representative values of 40\arcdeg\ and 70\arcdeg.

From the configurations generated with Starspotz, we find that the light curve can be explained by either a single 
high-latitude (70--90\arcdeg) spot subtending 1--20\% of the stellar surface or two spots at a large range of latitudes, 
with diameters from 3--20\arcdeg\ (up to 3\% of the area). The best fitting two-spot solutions depend highly on the 
adopted unspotted stellar brightness, as well as the spot darkness, and do not favor any particular latitude. We 
attempted to add three spots, but this did not improve the $\chi^2$ fit.

We have overlaid one of the best fitting models for HD~31305 on its light curve in Fig.~10. Deviations at the 0.5\% 
level, as well as a reduced $\chi^2$ value larger than 1100, confirm that there remains significant aperiodic variation 
in the data.

\subsection{AB Aurigae}

\begin{figure*}[!t]
\begin{center}
\plottwo{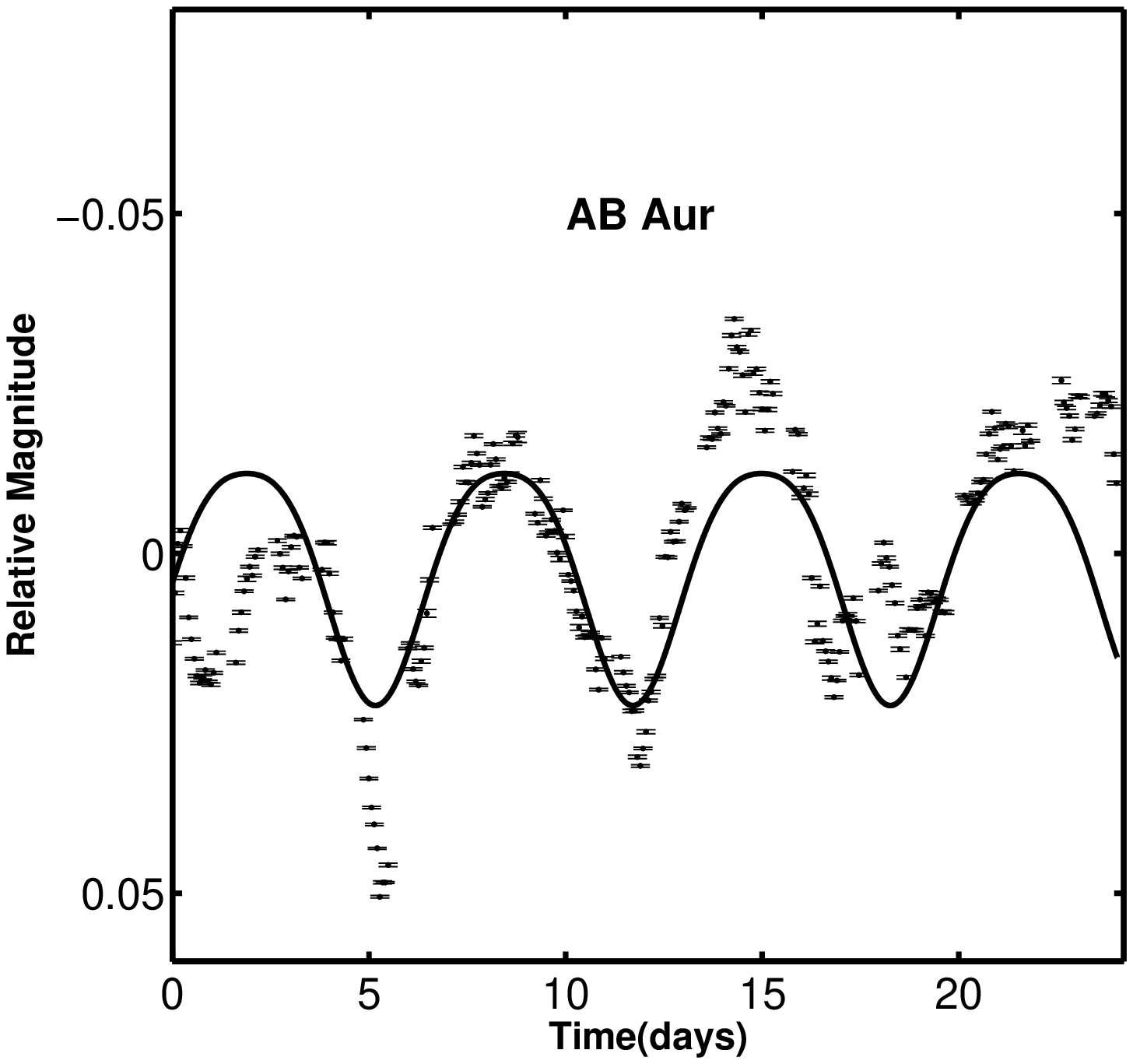}{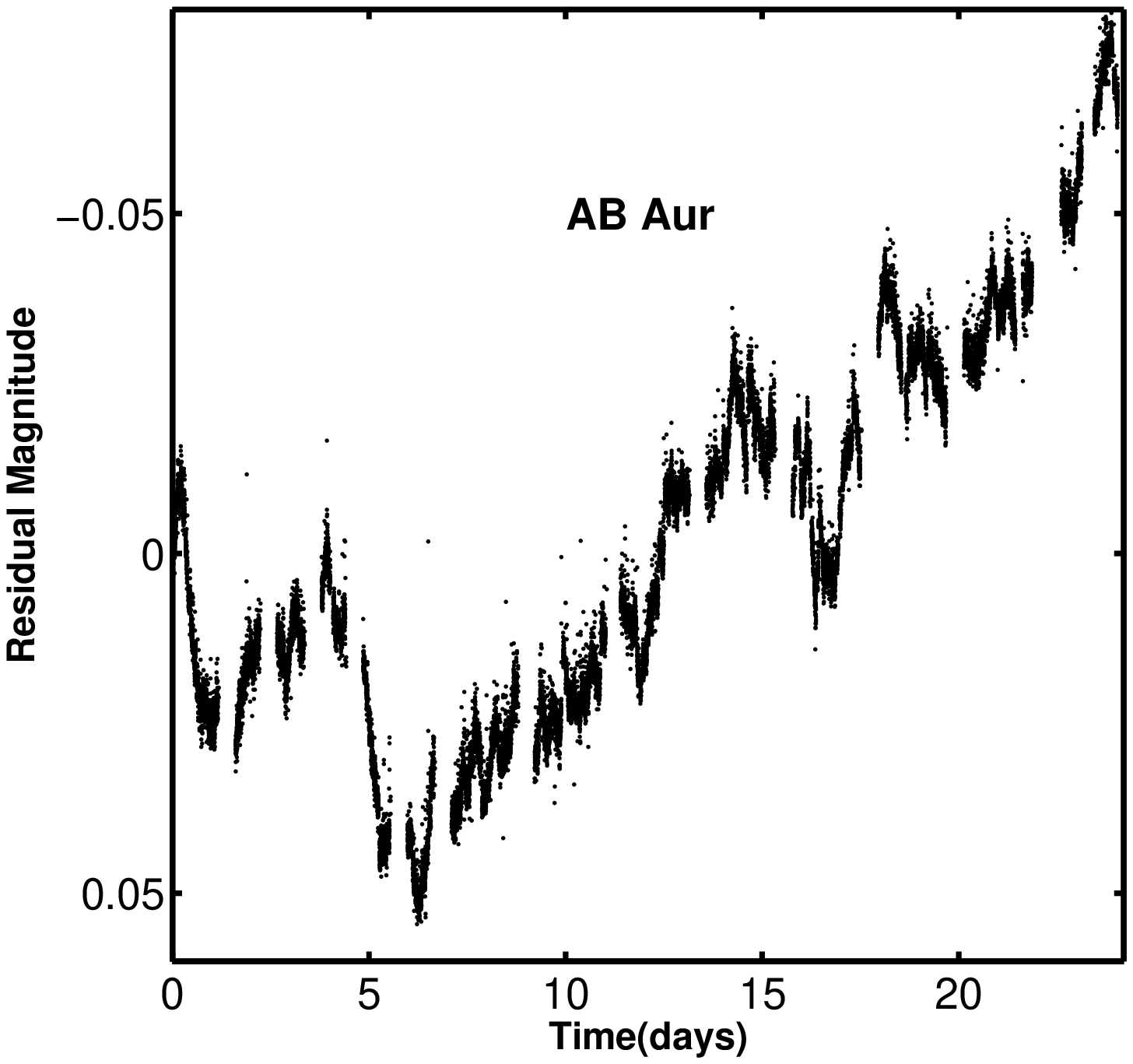}   
\end{center}
\caption[]{Left: Lightcurve of AB~Aur, binned on the {\em MOST} orbital timescale (points with uncertainties),
compared to one of the best fitting spot models (solid curve).
Right: Unbinned light curve residuals after model subtraction. Linear trends have been removed from the left light curve,
but not the right.}
\label{}
\end{figure*}

AB Aurigae is a Herbig Ae star with spectral type A0.  A large body of observations has 
indicated that it is encircled by an optically thick, gas dominated disk from 0.24 AU to 300 AU 
\citep[e.g.,][]{2008ApJ...689..513T,2008ApJ...679.1574O}.  This disk is viewed almost face-on, at an inclination 
angle between 12 and 35\arcdeg\ 
\citep{2005A&A...443..945P,2004ApJ...605L..53F,2005ApJ...622L.133C,2006ApJ...653.1353M}.

Previous photometric studies of AB~Aur uncovered periodic variability on short timescales from 0.5--1.8~d (Table 2). Our 
analysis found a much longer period (6.59~d), with substantial aperiodic light curve components on both longer and 
shorter timescales. Evidence that this is not the star's rotation period comes from the reported spectroscopic rotation 
velocity of $v$sin$i$=80$\pm$5~km~s$^{-1}$ \citep{1993A&AS..101..629B} and larger, with a radius of 2.0--2.5~$R_\odot$, 
assuming $L$=38--47 $L_\odot$ from \citet{2003ApJ...590..357D} and \citet{2008ApJ...689..513T}. These values are 
inconsistent with such a long period; even if $i$ were as large as 90\arcdeg, the photometric period would have to be 
less than 1.6 days.

Consequently, the variability observed in AB~Aur is not particularly amenable to spot model fitting. For completeness, 
and in case any of the reported parameters are erroneous, we have proceeded nevertheless to model its light curve using 
Starspotz. To begin, we subtracted out two linear trends at days 0--5 and 5--25 in the light curve, so that models could 
be fit to the shorter timescale ($<$10~d) variability in AB~Aur. We adopted two values for $i$: 22\arcdeg\ and 
35\arcdeg, matching the disk inclination values presented by \citet{2005ApJ...622L.133C} and 
\citet{2009ApJ...707L.132P}. Assuming one mid-latitude (20--60\arcdeg) spot subtending 10--15\arcdeg\ in diameter, we 
could reproduce some of the major trends of the data (see Fig.\ 11), and with a large number of small spots, we also 
reproduced some of the smaller fluctuations.  The results were relatively insensitive to the limb darkening parameter 
and the spot darkness. As suspected, none of the spot coverage scenarios provides a particularly satisfactory model, 
with reduced $\chi^2$ values in excess of 3000. However, we cannot rule out that a spotted photosphere contributes to a 
portion of AB~Aurigae's variability. As discussed for HD~31305, the light curve behavior observed here is not readily 
understood in an A0 star.

\section{Summary and discussion}

We have presented high precision, high cadence light curves for four known Taurus-Auriga members as well as one 
newly suspected member, HD~31305. The stellar masses for these stars range from $\sim$0.5--2.5~$M_\odot$. While our 
sample is not large, the high quality of the data has enabled us to probe young star variability down to 
sub-millimagnitude levels and search for low-amplitude and/or short-timescale phenomena.

\subsection{Discussion of lightcurve analysis} 

Our investigation has resulted in the measurement of photometric periods for these objects, all of which display a 
single dominant periodicity. For cases in which the periodic behavior can be explained by starspots, we have modeled 
the possible configurations and found that one mid- to high-latitude spot provides a sufficient, but far from 
perfect, explanation for the periodic variability. This picture is largely consistent with the results of 
\citet{2011MNRAS.415.1119S}, who carried out spot modeling of several variable young G and K stars. Among the 
periodic stars of their sample, they generally find a high latitude or polar spot (often large), and sometimes an 
additional smaller spot closer to the equator.

The main difference between our results and theirs is that we are unable to account for the full light curve 
behavior in our periodic objects using spot models alone. Additional aperiodic variability appears to be prevalent 
in the higher mass stars of our sample (AB~Aur, SU~Aur, and HD~31305) at the 0.1--1\% level, based on the flicker 
noise profiles in their periodograms and the substantial residuals after subtraction of best-fit spot models. The 
variability patterns among these three objects is nevertheless distinct, with differing characteristic time 
scales found by our autocorrelation analysis (Section 3.3). Investigation of a larger sample of young variable stars may 
shed light on how these differences arise; we suspect that the inclination of surrounding disks and accretion 
columns plays a substantial role in the properties of observed variability. In the case of SU~Aur, the dust is 
viewed nearly edge on \citep{2002ApJ...566.1124A,2005MNRAS.358..671K}, while for AB~Aur, it is close to face on 
\citep{2006ApJ...653.1353M}. Another nearly face-on object, TW~Hya, was shown to by \citet{2011MNRAS.410.2725S} to 
exhibit intermittent and variable-period oscillations mixed with more stochastic variability. MHD simulations of 
disk accretion by \citet{2008ApJ...673L.171R} have shown that there is a regime of suitably high accretion rates for 
which the flow of material becomes unstable, and the resulting light curve is stochastic.

The accretion rate of SU~Aur has been measured at a moderate 0.5--$0.6\times 10^{-8} M_\odot$~yr$^{-1}$ 
\citep{2004AJ....128.1294C}. Accretion rates reported for AB~Aur, on the other hand, vary widely, from relatively 
high values of 1.4 and 3$\times 10^{-7} M_\odot$~yr$^{-1}$ \citep{2006A&A...459..837G,2007ApJ...659..685B} to 
$7\times 10^{-8} M_\odot$~yr$^{-1}$ and lower based on more indirect methods 
\citep{1993A&AS..101..629B,2007A&A...468..541T}. It is therefore difficult to surmise whether how the variability 
observed in our target stars is connected with accretion properties.

%Magnetic Accretion and Photopolarimetric Variability in Classical T Tauri Stars- Stassun, Wood

Finally, we note that this work has resulted in the measurement of photometric rotation periods for {\em both} 
members of the binary V397~Aur. One period is close to but not exactly twice the other, suggesting that a tidal 
resonance may be operating in this system. The detection of multiple periods in a combined light curve also 
highlights the possibility that more close young binaries could be detected in this way without the need for 
extensive radial velocity data. For binaries in which the two rotation periods are similar, care must also be paid 
to distinguish the scenario from differential rotation.

% check for V397-- dynamical mass estimates?: Steffen, A. T. et al. A dynamical mass constraint for Pre-Main-Sequence 
% evolutionary tracks: the binary NTT 045251+3016. Astron. J., 122, 997-1006 (2001)

\subsection{HD 31305: An unusual variable young A star?}

As discussed in Section 4.3, the {\em MOST} light curve of HD~31305 resembles that of an active young star. Only recent 
analysis (see below) has hinted that it may be a member of the few-Myr-old Taurus-Auriga star-forming complex. All prior 
literature has assumed it to be a background field star, and in some cases, used it as a photometric comparison star 
\citep[e.g.,][]{1987AJ.....94..137H,2003ApJ...590..357D}!

A clue to its potential youth status is the infrared excess detected by IRAS (identifier 04526+3015), implying that 
HD~31305 is encircled by dust at $\sim$390~K \citep{1992A&AS...96..625O}. Recent observations from the WISE mission 
\citep{2011ApJS..196....4R} corroborate the presence of infrared excess, although the star was listed as a background 
object due to the absence of Taurus-Auriga membership information. It star has also been reported to flare in the X-ray 
\citep{2007A&A...468..477A} with a peak temperature of 8.6~keV \citep{2007A&A...471..951F}. These properties are largely 
inconsistent with those of A-type main sequence stars, which are weak X-ray emitters \citep{2003AdSpR..32..917L}. 
Additional evidence from Mooley et al.\ (2013, in preparation) bolsters the idea of a young age for HD~31305. They report a 
spectroscopic distance (174$\pm$11 pc) and proper motion consistent with Taurus membership at the 80\% confidence level. 
Their assembled SED from optical through mid-infrared wavelengths displays indications of a 10~$\mu$m silicate feature, 
and a model fit implies a temperature of $\sim$350~K at the inner edge of the dust region, consistent with the value 
suggested by \citet{1992A&AS...96..625O}. However, their low-resolution spectrum lacks emission lines, suggesting that 
the disk is not actively accreting. \citet{1997ApJ...475L..41M} pointed out that infrared excesses in HD~31305 and other 
A-type stars may be consistent with circumstellar dust shells. We alternatively suggest that it is weak enough to 
implicate a debris disk, since the ratio of infrared to stellar luminosity in the SED assembled by Mooley is 
$\sim$1:200. This scenario would not be inconsistent with an A star that is significantly older than $\sim$10~Myr, 
rather than the 3~Myr age of Taurus. Therefore, we cannot rule out that HD~31305 is a main sequence star, although its 
light curve behavior remains atypical and difficult to explain.

The variability itself has features in common with the other stars in our sample such as AB~Aur, specifically a 
light curve including both periodic and aperiodic components. To our knowledge, such a mixture of behavior lacks 
a mechanism among main sequence stars. The most common explanation for non-eclipsing periodic variability-- 
magnetic surface spots-- is difficult to invoke for A stars since they
are not believed to host substantial magnetic 
fields. The convective/radiative boundary vanishes for effective temperatures above $\sim$6000~K, implying the 
absence of a solar-like dynamo in these stars. Indeed, no spotted stars earlier than spectral type F8 ($T_{\rm 
eff}\sim$6300) have been observed. An exception is the class of chemically peculiar Ap and Bp stars, which retain 
magnetic fields left over as fossils from the formation process or produced by a turbulent dynamo in the core 
(Moss 1972). Objects in this class also exhibit large-scale chemical inhomogeneities on their surfaces, which 
are known to induce variability in the light curves of some Ap and Bp stars.

Although observational evidence for spots on high-mass stars remains weak, periodic variability in B-type main 
sequence stars attributed to rotational modulation has been detected in high-precision photometry from the {\em CoRoT} 
satellite \citep{2011A&A...536A..82D,2011A&A...528A.123P} as well as with ground-based data 
\citep{2001A&A...366..121B,2001A&A...380..177B,2004A&A...413..273B}. The periods of these B stars are between 0.3 
and 2 days, while their amplitudes are 1--2\% for the ground-based objects and a much lower $\sim$0.1--1 mmag for 
the objects observed with the high-precision {\em CoRoT} instrument. Perhaps more relevant to the question of HD~31305 are 
time series photometry from the Kepler mission indicating that up to 8\% of main sequence A stars display spot 
activity \citep{2011MNRAS.415.1691B}.  Notably, though, all of these detections were at an amplitude level well 
below 0.5 mmag (see their Fig.\ 3)-- a factor of 30 lower than the variability we detect in HD~31305. The 
possibility of spots on hot, non-peculiar stars is nevertheless supported by recent detections of weak magnetic 
fields in a handful of main sequence A stars including Vega \citep{2009A&A...500L..41L,2010A&A...523A..41P} and 
Sirius \citep{2011A&A...532L..13P}, as well as several hotter stars \citep{2011ASPC..449..275A}. While the paradigm 
of limited magnetic activity in these objects may be weakening, the variability characteristics associated with them 
still does not appear to be a good match for the behavior we have observed in HD~31305.

In contrast, the younger Herbig Ae/Be stars are enveloped in a complex and dynamic environment including 
circumstellar disks, inflowing material, and possible outflows, which might provide an explanation for the aperiodic 
components of HD~31305's light curve. These objects may also possess short-lived magnetic fields associated with the 
protostellar collapse and accretion processes. Such fields have been invoked to explain prominent X-ray emission 
observed in Herbig Ae stars \citep{2005ApJ...628..811S} and could be strong enough to generate surface spots 
responsible for the observed periodic variability. However, an analysis of the light curves of 230 Herbig Ae/Be 
stars \citep{1999AJ....118.1043H} revealed that while nearly all objects were variable, few displayed 
periodicities, on timescales up to 30 days, with amplitudes greater than several percent.
An additional possibility involving magnetic field generated variability is that we are 
observing photometrically not the rotation of the stellar surface, but the passage of dust clouds in the inner disk. 
The location of material orbiting at a Keplerian rotation period of 3.0 days around a young A0 star with an assumed 
mass of 2.4~$M_\odot$ is $\sim$0.05~AU. This is an order of magnitude closer to the star than the location implied 
by the 350--390~K dust temperature estimates of \citet{1992A&AS...96..625O} and Mooley et al.\ (2013, in preparation). 
Therefore, for the variability to be associated with a disk process, we speculate that copious material would have to be 
spiraling in toward the host star. Without further information about the nature of HD~31305's inclination and 
infrared excess, it is difficult to distinguish this scenario from periodic variability generated at the stellar 
surface.

A final option is that HD~31305 itself is not variable, but rather has a spotted lower mass companion contributing to the 
overall flux variations and producing the X-ray emission, as proposed by \citet{2007A&A...468..477A}. We checked the Keck 
Observatory Archive and discovered an exposure of this star taken with the NIRC2 adaptive optics imaging system as part 
of engineering tests. The $K_{\rm p}$ image reveals that there is indeed a companion lying 0.52\arcsec\ away, at a 
position angle of $\sim$128.5\arcdeg. We evaluate the possibility that this object is the source of variability in our 
{\em MOST} observations by considering its brightness and inferring the implied variability amplitude. Based on the NIRC2 
image we estimate the companion to be 1.3 mag fainter than the primary A0 star at the $K_{\rm p}$ band. Since a 
0.5\arcsec\ separation indicates a significant probability of association, we consider the case that the two stars are 
bound members of the $\sim$3~Myr Taurus-Auriga association. The TA-DA tool \citep{2012AJ....144..176D} enables us to 
compute isochrones, infer a mass or temperature for the companion, and calculate its expected brightness in the {\em 
MOST} band, assuming no extinction. Inputting $\Delta K_{\rm p}$=1.3, we find an expected companion temperature of 
$\sim$4900~K, or K2 spectral type. Integrating a K2 model spectrum over the {\em MOST} band from 350--750~nm, we predict 
the companion to be 3.7 mag fainter at optical wavelengths.  Because of {\em MOST}'s 3\arcsec\ pixel size, the 
light curve of HD~31305 includes all of the flux from this neighboring star. Thus the lower mass star contributes just over 3\% 
of the total flux. To cause the observed $\sim$1.5\% variations, the companion's brightness would have to vary by 47\%, 
or 94\% peak to peak. This value is at the upper end (i.e., $<$1\% level) of the stellar activity amplitude distribution 
for similar temperature stars \citep[see Fig.\ 3 of][]{2010ApJ...713L.155B}, and is exceedingly rare among K-type periodic 
pre main sequence stars \citep[e.g.,][]{1999AJ....117.2941S,2004AJ....127.1602C}. Furthermore, it is difficult to conceive
of a spot large enough to cause such a substantial variability amplitude. Thus a spotted companion does not 
provide a likely explanation for the light curve behavior in our {\em MOST} time series.

We conclude that the variability may come from the A0 star itself. Given the lack of previous 
detections of periodic variability in main-sequence A stars, we believe the most likely scenario for HD~31305 is that 
it is relatively young and displays a heretofore undetected type of variability, which we tentatively associate with 
its dusty excess or a magnetic field.

\subsection{The origin of periodic variability in SU Aurigae}

Our {\em MOST} observations SU~Aur have resulted in the first detection of clear periodic behavior in this star, at a 
period of 2.66 days. Although SU~Aur has been monitored for years, analysis of its light curve has focused on the 
prominent fading events that led to its classification as an UXOR object. There have been a number of periodicity claims 
(see Table 2), but most were based on spectroscopic or X-ray data, and many of the periods were noted to have large 
uncertainties or marginal significance. We attempted to recover the 2.66~d periodicity in publicly available data from 
the T Tauri photometry database of \citet{1994AJ....108.1906H} but failed to detect any peaks at the corresponding 
frequency (0.38 d$^{-1}$) in the periodogram. It is conceivable that the combination of sparse sampling and systematic 
dimming events in that dataset may have masked transitory periodic behavior. Higher cadence monitoring is necessary to 
determine the fraction of time for which periodic variability manifests in SU~Aur's light curve.

As pointed out in Section 4.2, the $v$sin$i$ value of SU~Aur is $\sim$66.2~km~s$^{-1}$, with an uncertainty of 
$\sim$4.6~km~s$^{-1}$ \citep{1986ApJ...309..275H}. If the stellar inclination is in line with that of the disk 
\citep[62\arcdeg;][]{2002ApJ...566.1124A}, then the expected equatorial velocity is 75$\pm$7 km~s$^{-1}$. In this 
case, the only way to infer an rotation period as long as the one we have measured (2.66~d) is if the stellar radius 
is at least $\sim$3.6~$R_\odot$. If we discount the disk inclination and raise $i$ to 90\arcdeg, then the radius may 
be as low as 3.2~$R_\odot$.

These values stand in contrast to the $R=2.75\pm 0.25$ derived by \citet{2003ApJ...590..357D} from spectral energy 
distribution (SED) fitting. \citet{1989ApJ...339..455C} reported a much larger radius (3.6~$R_\odot$), which we trace to 
the larger value of the bolometric luminosity derived by them ($\sim$13~$L_\odot$ vs. $\sim$6~$L_\odot$ from 
\citet{2003ApJ...590..357D}). The stellar luminosity of SU~Aur is difficult to determine accurately because of the disk 
contribution to its flux at near-infrared and infrared wavelengths. We have derived our own luminosity values using 
photometry listed in the SIMBAD database, along with reddening corrections from a fit to the SED (spectral type G2) and 
band-dependent bolometric corrections. Depending on the bands used, the stellar luminosity could lie anywhere between 7 
and 30 $L_\odot$. We favor the lower luminosities of 7--9~$L_\odot$, as these result from SED fits to the $U$, $B$, and 
$V$ bands only and exclude contaminating emission from the disk at longer wavelengths. Taking these systematics into 
account, we estimate the radius of SU~Aur to be between 2.9 and 3.2~$R_\odot$. While we cannot rule out a higher 
luminosity and hence larger radius, the above estimates of spectroscopic rotation velocity and the 2.66~d period suggest 
that these values are unphysical.

The inconsistency between our measured periodicity and the inferred rotation rate of SU~Aur leads us to 
believe that we may be witnessing the motion of a dust cloud or hot spot connected with this star's inner disk. On 
the presumption that we are not observing starspots, but rather observing material in Keplerian orbit about the 
star, we can derive its location by adopting the mass of 2.0$\pm$0.1~$M_\odot$ derived by 
\citet{2003ApJ...590..357D}. We find an orbital distance of 7.1$\times$10$^{11}$~cm, or $\sim$0.05~AU. Intriguingly, 
this is very close to the value of the inner disk edge reported by \citet{2002ApJ...566.1124A} based on 
interferometric observations of SU~Aur (0.05--0.08~AU). We therefore conclude that the source of periodic 
variability observed during our {\em MOST} observations could be a structural or thermal feature on the inner disk 
edge, or possibly a discrete cloud of material orbiting along magnetic field lines \citep[e.g.,][]{1997ApJ...486..397U}
just interior to the disk.

Measurement of the rotation period of the inner disk also has implications for the angular momentum evolution of young 
stars, many of which are believed to be magnetically locked to their disks for at least a few Myr \citep[see][for a 
recent summary]{2012ApJ...756...68C}. Our inference of a star rotating significantly faster ($P\sim$1.86 days, based on 
the $v$sin$i$ and $i$ values adopted above) than the inner disk confronts this idea. The magnetic field lines connecting 
these two regions would likely be highly distorted by this velocity shear, and the properties of accretion flows could 
be quite different from those predicted by models involving corotating disks \citep{2008ApJ...673L.171R}. Alternatively, 
we could be observing disk material that is located farther out than the inner edge, and thus beyond the corotation 
radius.

%On late November and early December 2006, NARVAL and ESPaDOnS simultaneously observed SU Auriga, a newly-born star 
%weighting about twice as much as the Sun and located at a distance of about 450 light years from us. With an age of only a 
%few million years (as opposed to 5 billion years for the Sun), SU Aur is still in its extreme infancy.
%"Recent observations by Nadalin et al. (2000) have shown short time variability in the B-band magnitude that they attribute to 
%protoplanetary materials orbiting in the circumstellar disk."

\subsection{Implications for future time series observations of young stars}

We have seen that full characterization of young star variability requires a combination of high photometric 
precision, short cadence, and sufficiently long time baseline that is challenging to arrange. The appearance of both 
periodic and aperiodic variability suggests that it may be difficult to infer rotation periods with sparsely sampled 
data, as is usually the case from the ground. The large range of rotation periods reported for some of our targets 
underlines this problem. 

In the cases of SU~Aur and HD~31305, we have further suggested that periodic variability may not represent the stellar 
rotation period at all, if it is instead tied to the surrounding circumstellar disk. We therefore highlight a need for 
more comparisons of $v$sin$i$ measurements with photometric periods detected in young stars with infrared excesses. We 
suggest that these measurements be carefully reviewed for stars displaying infrared excesses or spectroscopic accretion 
signatures suggestive of disks. Rotation period distributions for these objects typically have been derived from 
ground-based data, and in some clusters display bimodal structure \citep{Herbst02}. Periodicities associated with 
regions many stellar radii above the star's surface may contaminate rotation rate samples and thus bias our view of 
angular momentum evolution if not all young cluster stars are locked to their disks.

Different observational complications arise for aperiodically variable targets. Our timescale analysis has provided 
guidelines for the time sampling rate in photometric future campaigns. For objects with aperiodic light curve behavior, 
our results suggest that data need not be taken more frequently than every hour to characterize the 
stochastic components of variability in young stars at 1\%
precision. For higher signal-to-noise observations such as those
presented here, cadences as short as five minutes can be used to probe lower
amplitudes of the flicker noise. In addition, we find a wide range of coherence 
timescales characterizing the aperiodic variability, from 0.2 to 6.2 days.  However, our very small sample may not be representative of 
all young stars, especially those in the very low mass range below 0.5~$M_\odot$. In particular, accreting brown 
dwarfs may have shorter characteristic variability timescales, as measured by autocorrelation \citep[e.g.,][]{Cody11}. 
We thus encourage further high-precision, well sampled long baseline time series observations of young cluster members 
to expand the size and diversity of the variability dataset. Clearly, multi-wavelength data as well as high-resolution 
spectroscopy will provide much-needed feedback for model development.

\acknowledgements{Thanks to Kunal Mooley, Konstanze Zwintz, Slavek Rucinski, Scott Gregory, and Evelyn 
Alecian for helpful discussions. These observations were obtained under NASA grant NNX09AH27G.}

\bibliographystyle{apj}
\bibliography{ms}

\end{document}